\documentclass[aps,prl,twocolumn,groupedaddress,showpacs,superscriptaddress,amssymb,amsmath]{revtex4-1}
\usepackage{physics}
\usepackage{verbatim}
\usepackage{graphicx}
\usepackage{amsmath}
\usepackage{amssymb}
\usepackage{amsfonts}
\usepackage{color}
\usepackage{bm}        % for math
\usepackage{comment}
\usepackage{placeins}
\usepackage[mathscr]{eucal}
\usepackage[colorlinks, linkcolor=red,citecolor=blue,urlcolor=blue]{hyperref}
\usepackage[normalem]{ulem}
\usepackage[all]{hypcap}
\usepackage{multirow}
\usepackage[dvipsnames,usenames,table]{xcolor}
\usepackage{dcolumn}
\usepackage{hyperref}
\usepackage{bm}
\usepackage{epsf}
\usepackage{subfigure}
\usepackage{amsfonts}
\usepackage{epstopdf}%
\newcommand{\be}{\begin{equation}}
\newcommand{\ee}{\end{equation}}
\newcommand{\bea}{\begin{eqnarray}}
\newcommand{\eea}{\end{eqnarray}}
\definecolor{mblue}{rgb}{0.0,0.45,0.74}
\usepackage{hyperref}
\usepackage{amsfonts}
\usepackage{tikz}
\usepackage{pgfplots}
\DeclareRobustCommand{\rdot}{\raisebox{1pt}{\tikz{\draw[gray,dotted,line width=2pt](0,0) -- (3.5mm,0);}}}
\DeclareRobustCommand{\sblack}{\raisebox{1.5pt}{\tikz{\draw[mblue,solid, line width=1pt](0,0) -- (5mm,0);}}}
\DeclareRobustCommand{\dablack}{\raisebox{1.5pt}{\tikz{\draw[gray,dashed,line width=1pt](0,0) -- (5mm,0);}}}
\DeclareRobustCommand{\dblack}{\raisebox{1.5pt}{\tikz{\draw[mblue,dotted,line width=1.5pt](0,0) -- (5.3mm,0);}}}

\begin{document}

\title{Strong system-bath coupling reshapes characteristics of quantum thermal machines}
%absorption refrigerators}

%\title{Strong system-bath interaction reshapes  the performance of quantum absorption refrigerators}

\author{Felix Ivander}
\altaffiliation{These authors contributed equally to this work}
\affiliation{Chemical Physics Theory Group, 
Department of Chemistry and Centre for Quantum Information and Quantum Control,
University of Toronto, 80 Saint George St., Toronto, Ontario, M5S 3H6, Canada}
\author{Nicholas Anto-Sztrikacs}
\altaffiliation{These authors contributed equally to this work}
\affiliation{Department of Physics, 60 Saint George St., University of Toronto, Toronto, Ontario, Canada M5S 1A7}
\author{Dvira Segal}
\affiliation{Chemical Physics Theory Group, 
Department of Chemistry and Centre for Quantum Information and Quantum Control,
University of Toronto, 80 Saint George St., Toronto, Ontario, M5S 3H6, Canada}
\affiliation{Department of Physics, 60 Saint George St., University of Toronto, Toronto, Ontario, Canada M5S 1A7}
\email{dvira.segal@utoronto.ca}

\date{\today}

\begin{abstract}
We study the performance of quantum absorption refrigerators, 
paradigmatic autonomous quantum thermal machines,
and reveal central impacts of strong couplings between the working system and the thermal baths.
% on the  cooling performance.
Using the reaction coordinate quantum master equation method, 
which treats system-bath interactions beyond weak coupling, we discover
that {\it reshaping} of the window of performance is
a central outcome of strong system-bath couplings. 
This alteration of the cooling window stems from the dominant role of
parameter renormalization at strong couplings.
We further show that strong coupling admits  %(beyond second order)
direct transport pathways between the thermal reservoirs.
Such beyond-second-order transport mechanisms are typically detrimental 
to the performance of quantum thermal machines.
%
%Our comprehensive study, exposing the reshaping of the operation window at strong coupling,
Our study reveals that it is inadequate to claim for 
either a suppression or an enhancement 
of the cooling performance at strong coupling---when analyzed against a single parameter and 
in a limited domain. Rather, a comprehensive approach should be adopted so as
to uncover the reshaping of the operation window.
%strong coupling is to trasnform-reshape the performance window.

%
%We conclude that strong system-bath coupling offers no fundamental 
%advantage to the operation of autonomous quantum thermal machines.
%
%
\end{abstract}

\maketitle

%========================================================================
{\it Introduction.--}
%\section{Introduction}
%quantum thermodynamics -> heat machines 
The development of classical thermodynamics during the 19th century was directly 
tied to its application in building macroscopic heat engines. 
Fast forward to present times, the field of quantum thermodynamics \cite{Anders_2016,Goold_2016,Kosloff_2013} 
is similarly advanced by analyzing concrete machines such as the quantum Otto cycle or quantum absorption refrigerators.
% experiments
An early theoretical exploration of a small-scale quantum device was done by 
Scovil and Schulz-DuBois \cite{maser1959} who showed that a 3-level system could 
operate as a heat engine (maser). 
Nowadays, miniaturized thermal machines are experimentally realized from systems as small as a few 
trapped ions \cite{Maslennikov2019}, a single atom \cite{Schmidt-Kaller2016}, and
electron spins \cite{Schmidt-Kaller_2019}.    
%
%introducing QTMs and discussing various ressources for them
Quantum thermal machines are distinct from their classical counterparts by relying on nonclassical
effects such as quantum coherences \cite{Kosloff_2015,Streltsov2017,Kilgour2018,Binder_2020,Petrucionne_2021}, 
correlations \cite{Brunner_2014,Bresque2021}, 
measurements \cite{Elouard_2017,Buffoni_2019,Bresque2021} and statistics \cite{Deffner}, which 
may serve as a resource to enhance their performance beyond their classical analogues; 
studies cited here are representative of a growing literature.

%Strong coupling 
Though not inherently a quantum mechanical effect, strong system-reservoir interactions 
are a common characteristic of quantum thermal devices due to the typical small scale of these machines
and the fact that phase-coherent dynamics retains boundary effects. 
%The interaction energy associated with the system-bath coupling is typically ignored in classical thermodynamics given its minute magnitude compared to the system's energy.
%This immediately brings up the question: 
What is the impact of strong system-bath coupling on the performance of quantum thermal machines? 
Can we harness strong couplings to realize regimes of operation that are inaccessible 
at weak coupling?
These questions were examined in different models, bringing model-dependent, sometimes conflicting answers 
% conflicting
%Understanding the impact of strong coupling on the performance of quantum thermal machine is an active area of research
\cite{Strasberg_2016, Tanimura_2016,Eisert_2018, Newman_2017,Mu_2017, Hava_2019,Nitzan18,Newman_2020, McConnell_2021, Mu_2017, Wiedmann_2020, Zhang,Goold_2020,Tanimura20,JunjieS,Galperin}. 
On the one hand, strong system-bath coupling may open  transport pathways 
that are inaccessible at weak coupling (e.g. by admitting high order processes 
and cooperative effects) \cite{Mu_2017,Hava_2019}.
% thus going beyond selection rules)
On the other hand, strong coupling between a system and its environment can 
suppress currents due to particle-dressing (polaron) physics \cite{Nicolin_2011,Hava_2019,Nick_2021}.
%=-==========================
% Figure 1
\label{Model}
\begin{figure}[tb]
\centering
\includegraphics[width=0.75\columnwidth]{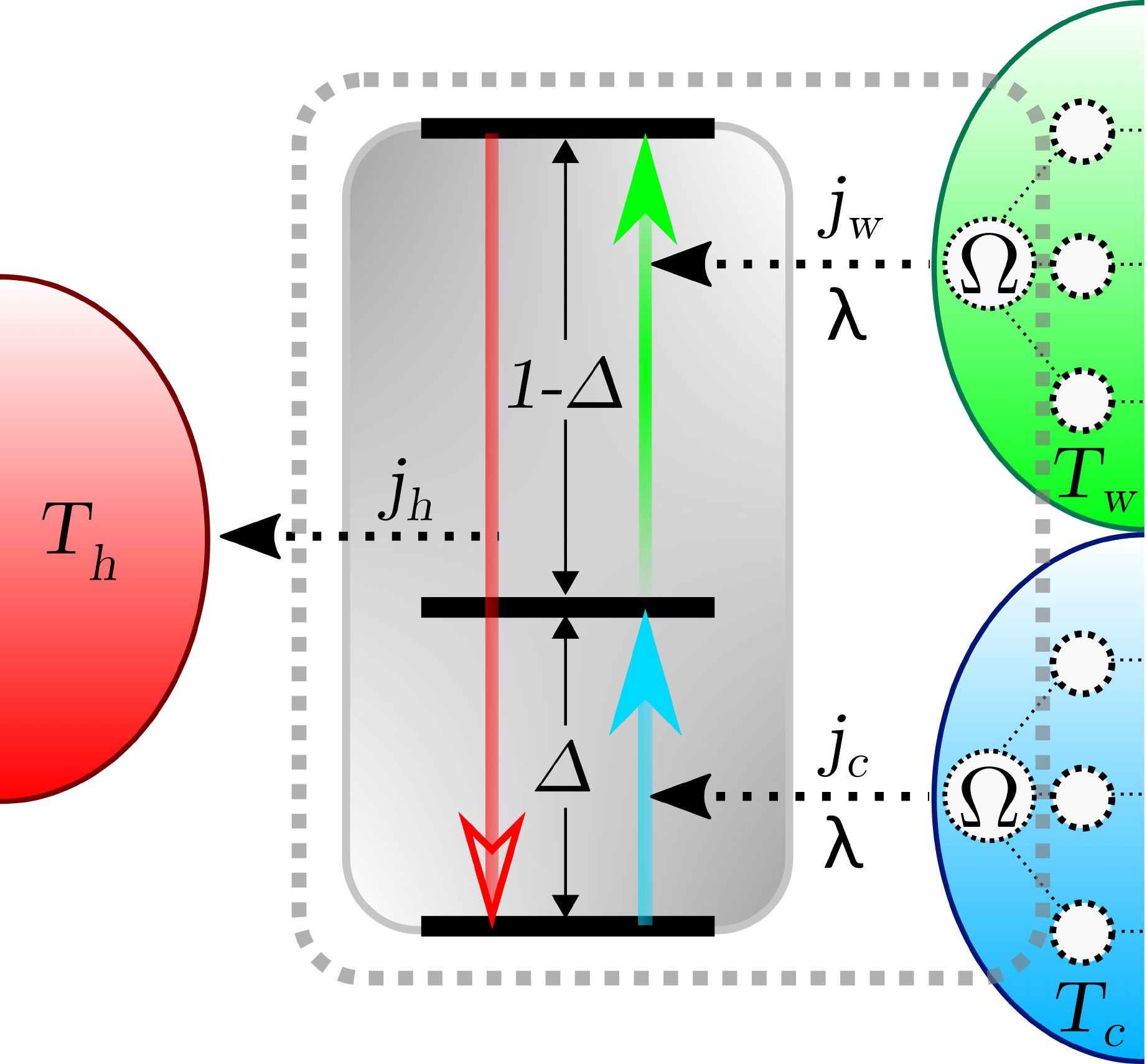} %{Figures_QAR/Truefinal/diag.png}
\caption{Scheme of a 3-level QAR  as investigated in this study.
%with energy spacings $\Delta$ and $1-\Delta$. 
%In the weak coupling process, 
%Here, the dashed arrows denote a sequential energy transfer process induced by one of the baths to generate excitation/relaxtion between two energy levels.
The thick blue, green and red arrows represent transitions inducted by the cold, work and hot baths,
respectively, exemplifying a cooling process.
The dashed line (\rdot) that encircles the system and part of the baths represents the 
repartitioning of the system and environment with the
incorporation of collective bath coordinates
%---from the cold and work baths---
into the system.}
    \label{fig:Fig1}
\end{figure}
%================================

In this work, our goal is to provide a deep understanding over the impact of strong 
system-bath coupling on the performance of continuous-autonomous thermal machines, specifically, 
the paradigmatic 3-level model of a quantum absorption refrigerator (QAR) \cite{maser1959, KosloffQAR,
Correa_2015,Mitchison}.  
This thermal machine extracts heat from a cold reservoir and transfers 
it to a hot bath by utilizing heat absorbed from an even hotter, ``work" reservoir.  
QARs can be constructed from a single qubit \cite{Mu_2017}, 
interacting qubits \cite{Correa14,Mitchison_2016}, 
or multi-level quantum systems \cite{Kilgour2018}. 
In this study, we concentrate on the 3-level model of QARs \cite{KosloffQAR,Correa14,Mitchison_2016}, 
depicted in Fig. \ref{fig:Fig1}.
Recent studies examined different aspects of this machine, %related to the performance of QARs,
such as the impact of dissipation and  current leaks \cite{Correa_2015,HavaM}, 
the role of internal coherences \cite{Kilgour2018}, 
current fluctuations \cite{Segal2018,Junjie_2021}, and topology \cite{Gonzalez_2017}. 
Furthermore, experimental implementations have been proposed \cite{Hofer_2016} 
and realized \cite{Maslennikov2019}. 
While the question over the impact of strong system-bath 
coupling on the performance of QARs has been examined in 
different studies \cite{Strasberg_2016,Tanimura_2016,Mu_2017, Hava_2019}, %XXX
these sporadic results do not allow a comprehensive understanding. 
Is strong system-bath coupling beneficial or detrimental to the cooling operation of QARs?

Here, we reveal that a prominent impact of strong system-bath coupling in quantum thermal machines
is the {\it reshaping} of the window of operation due to parameter renormalization.
Watching the cooling current in a limited region may lead to conflicting
conclusions as to whether it is enhanced or suppressed at strong coupling.
In contrast, by using the powerful---and economic---reaction coordinate (RC) quantum master equation (QME)
technique, which
captures system-bath interactions beyond the weak coupling regime 
\cite{Burghardt1,Burghardt2,Burghardt3,
NazirPRA14,Nazir16,Nazir18,Correa19,GernotF,Nick_2021,Camille}, 
we observe %expose the full role of strong system-bath interactions on performance, and observe
the alteration of the performance window due to strong coupling.
Besides reshaping the cooling window, we show that strong coupling opens direct bath-to-bath
transport pathways, typically detrimental for the performance of quantum thermal machines.

{\it Model.--} A quantum absorption refrigerator cools an 
environment (temperature $T_c$) by transferring its heat into a hot reservoir ($T_h$). 
This is achieved with an energy input from a ``work" bath ($T_w$), where $T_c<T_h<T_w$. 
The process is mediated by a three-level system ($i=1,2,3$) playing the role of the classical working fluid.
Transitions between energy levels of the quantum system are induced by the
different reservoirs, see Fig. \ref{fig:Fig1}. 
The Hamiltonian describing this setup is given by
\bea
\label{eq:QARHamiltonian}
&&\hat H_{QAR} = \sum_{i = 1}^3 \epsilon_i \ket{i}\bra{i} 
%\nonumber
 \\
 &&+ \sum_{\alpha=\{c,w,h\}}\sum_k \nu_{\alpha,k} \left( \hat c_{\alpha,k}^{\dagger} + 
\hat S_{\alpha}\frac{f_{\alpha,k}}{\nu_{\alpha,k}} \right)\left( \hat c_{\alpha,k} + \hat S_{\alpha}\frac{f_{\alpha,k}}{\nu_{\alpha,k}} \right). 
   \nonumber
\eea
Here, $\epsilon_i$ denotes the energy of the $i$th level of the QAR. 
%For simplicity, we choose the ground state to be the zero energy mark, i.e. $\theta_0 \equiv 0$. 
In what follows and without loss of generality, 
we set the total energy gap as $\epsilon_3-\epsilon_1=1$, and define the first gap by
$\Delta\equiv\epsilon_2-\epsilon_1$.
%Moreover, we label the energy of the excited states as $\Delta$ and $g$, respectively and set $g = 1-\Delta$. 
As for the baths, $\hat c_{\alpha,k}^{\dagger}$ ($\hat c_{\alpha,k}$) are 
bosonic creation (annihilation) operators of the $\alpha=c,h,w$ reservoir 
for a mode of frequency $\nu_{\alpha,k}$. 
The baths' harmonic oscillators couple to different transitions in the system 
with strength $f_{\alpha,k}$, assumed a real number. 
The system operators $\hat S_{\alpha}$ couple to the different baths.
The effects of these interactions are captured by the spectral density functions,
$J_{\alpha}(\omega) = \sum_{k}f_{\alpha,k}^2\delta(\omega-\nu_{\alpha,k})$.
Setting the allowed transitions as
%\bea
    $\hat S_c = \ket{1}\bra{2}$ + h.c,
 %   \nonumber \\
    $\hat S_h = \ket{1}\bra{3}$ + h.c, 
 %   \\
    $\hat S_w = \ket{2}\bra{3}$ + h.c
%\eea
allows for the maximal (Carnot) efficiency to be reached at weak coupling at a certain point
\cite{KosloffQAR,Correa_2015,Kilgour2018}.
For reasons explained shortly, 
the spectral density function %describing the interaction between the reservoir and the system 
is chosen to be a peaked Brownian oscillator function for the cold and work baths  ($\alpha=w,c$),
\bea
 \label{eq:Brownian}
 J_{\alpha}(\omega) = \frac{4 \omega \gamma_{\alpha} \Omega_{\alpha}^2 \lambda_{\alpha}^2}{(\omega^2 - \Omega_{\alpha}^2)^2 + (2\pi \gamma_{\alpha} \Omega_{\alpha} \omega)^2}.
\eea
Here, $\lambda_{\alpha}$ captures the system-environment coupling strength, and it is the central parameter of this work.
$\gamma_{\alpha}$ is a measure of the width of the spectral function, 
and $\Omega_{\alpha}$ is the frequency about which the spectral density is peaked. To contrast, the hot bath spectral density is chosen to be an Ohmic function
\bea
J_h(\omega) = \gamma_h \omega e^{-|\omega|/\Lambda_h},
\eea
with $\gamma_h$ dictating the coupling strength between the hot bath and the QAR and $\Lambda_h$ a high frequency cutoff.  

{\it Reaction coordinate mapping.--} 
To go beyond a weak coupling treatment, we extract
collective modes (``reaction coordinates") from the heat baths and include them within
the system's Hamiltonian, to redefine the system-environment boundary. 
For computational efficiency, we assume that the system is weakly coupled to the hot ($h$) bath, 
while allowing arbitrary coupling to the cold and work baths. 
We thus extract two RCs, from the cold and work baths, each.
It is straightforward to further include a collective mode from the hot bath. 
This exact mapping results in the following, RC-QAR Hamiltonian \cite{Nazir18,Nick_2021}
%
%\begin{widetext}
\bea
% \nonumber
&&\hat H_{RC-QAR} =  \sum_{i = 1}^3 \epsilon_i \ket{i}\bra{i} +
\sum_{\alpha=\{c,w\}} \Omega_{\alpha} \hat a_{\alpha}^\dagger \hat a_{\alpha} 
\nonumber\\
&&+\sum_{\alpha=\{c,w\}} \hat S_{\alpha} \lambda_{\alpha}(\hat a_{\alpha}^\dagger+\hat a_{\alpha}) 
+ \sum_{\alpha=\{c,w\}} \frac{g_{\alpha,k}^2}{\omega_{\alpha,k}} (\hat a_{\alpha}^{\dagger}+\hat a_{\alpha})^2
\nonumber \\
 && + \sum_{k} \nu_{h,k} \left( \hat c_{h,k}^{\dagger} + \hat S_h\frac{f_{h,k}}{\nu_{h,k}} \right)
\left( \hat c_{h,k} + \hat S_h\frac{f_{h,k}}{\nu_{h,k}} \right) 
\nonumber\\
&&+ \sum_{\alpha=\{c,w\}} (\hat a_{\alpha}^{\dagger}+\hat a_{\alpha})\sum_k g_{\alpha,k}
 (\hat b_{\alpha,k}^{\dagger}+\hat b_{\alpha,k}) 
%+ \sum_{\alpha} \frac{g_{\alpha,k}^2}{\omega_{\alpha,k}} (\hat a_{\alpha}^{\dagger}+\hat a_{\alpha})^2
 \nonumber\\
 &&+ \sum_{k} \nu_{h,k} \hat c_{h,k}^\dagger \hat c_{h,k}
 + \sum_{\alpha=\{c,w\};k} \omega_{\alpha,k} \hat b_{\alpha,k}^\dagger \hat b_{\alpha,k}.
%+ \sum_{k} \nu_{h,k} \hat c_{h,k}^\dagger \hat c_{h,k} 
    \label{eq:RC-QAR Hamiltonian}
\eea
%\end{widetext}
%
The first four terms constitute the {\it extended system} with
$\hat a_{c,w}^{\dagger}$ ($\hat a_{c,w}$) as bosonic creation (annihilation)
operators of two RCs with frequencies $\Omega_{c,w}$,
extracted from the cold and work environments, respectively.
The RCs are coupled to the 3-level system by $\lambda_{\alpha}$.
Note the quadratic term in the RC, which emerges after the mapping.
The third and fourth lines describe the coupling of the system to the hot bath, 
unaltered by the mapping, and the coupling of the RCs to the work and cold baths; 
$\hat b_{\alpha,k}^{\dagger}$ ($\hat b_{\alpha,k}$) are operators of the residual  
baths and $g_{\alpha,k}$ are coupling parameters that build the spectral density functions 
$J_{RC,\alpha}(\omega)=\sum_k g_{\alpha,k}^2\delta(\omega-\omega_{\alpha,k})$. %% DDDN
The last line includes the Hamiltonians of the original hot bath, and the 
residual cold and work reservoirs.

%As outlined in e.g. Refs. \cite{NazirPRA14,Nick_2021}, 
If the spectral density function for the original cold and work baths, Eq. (\ref{eq:QARHamiltonian}), is Brownian, 
Eq. (\ref{eq:Brownian}), then
the spectral density functions of the residual baths become Ohmic with an 
infinite high frequency cut-off,
%
%\bea
 $ J_{RC,\alpha}(\omega)=\gamma_{\alpha}\omega e^{-\abs{\omega}/\Lambda_{\alpha}}$
%\eea
%
\cite{NazirPRA14,Nick_2021}.
In the RC representation, $\gamma_{\alpha}$ is a dimensionless coupling constant of the 
RC to the residual bath---later assumed small to justify
a perturbative treatment for the extended system. 
Furthermore, $\Lambda_{\alpha}$ is the cutoff frequency of the $\alpha$th bath. 
%taken as a large parameter.
%
We perform the RC mapping on the work and cold baths only. 
Thus, the hot bath is assumed Ohmic from the start, 
while the cold and work residual baths are Ohmic only after the mapping.
For convenience, we assume a symmetry between the baths such that 
$\Omega_c = \Omega_w = \Omega$, $\lambda_c = \lambda_w = \lambda$ and 
$\gamma_{c,h,w} = \gamma$.

%==========================================================
%\section{Method}
%============================================
%\begin{figure}
%    \centering
%    \includegraphics[width=1\columnwidth]{Figures_QAR/Final/FIG3F.png}
%    \caption{(a). The first six eigenenrgies of  of the enlarged QAR-RC Hamiltonian. %$\hat H_{ES}^M$. 
%    (b) We focus on the lowest three levels, which roughly correspond to the RCs in their ground state. 
%   % Parameters are taken to be: $\Delta = 0.2$. 
%   %All other parameters are identical to those used in Figure \ref{fig: Fig2}.}
%  $\Omega=20$, $M$=.
%   }
%    \label{fig:Eigenspectrum}
%\end{figure}
%==========================

{\it Method.}
%The effectiveness of the RC-QME method stems from its simplicity in computing open system dynamics 
%and steady state properties. 
Due to the assumption of weak coupling between the {\it extended system} and the 
residual baths (small $\gamma)$, 
one can employ standard perturbative quantum master equation tools to study transport
behavior in the RC-QAR model, a method referred to as the RC-QME  \cite{NazirPRA14,Nazir16,Nick_2021}. 
The procedure involves three steps.
(i) We truncate the two RCs harmonic oscillators to include $M$ levels each.
Working in the occupation basis, the resulting $3\times M^2$-dimension extended system's
Hamiltonian is denoted by $\hat H^M_{ES}$.
We note that the computational complexity of the RC-QME method rapidly
increases with the dimensionality of the extended system as $(3 \times M^2)^4$.
The value of $M$ is decided on by converging simulations of
the desired expectation value, see \cite{SuppI}. 
(ii) We diagonalize the extended system $\hat H^M_{ES} \to $ $\hat H_{ES}^D$, and
represent the system-bath coupling operators in the energy eigenbasis \cite{SuppI}.
(iii) We use the Redfield equation to study steady state transport 
of the $3M^2$-dimension extended system.
This amounts to the assumptions that the RC is weakly coupled to the residual environments, the
dynamics is Markovian, and the initial state of the open system 
is a product state of the extended system and baths, the latter prepared in canonical states.
Formally, the Redfield equation is given by
%
%\bea
$\dot \rho_{ES}(t)=-i[\hat H_{ES}^{D},\rho_{ES}] + \sum_{\alpha}{\cal D}_{\alpha}(\rho_{ES})$.  % + D_h(\rho_{ES}) + D_c(\rho_{ES})$,
%\eea
%
The dissipators ${\cal D}_{w,h,c}(\rho_{ES})$ are additive given the weak coupling approximation of the RCs to their residual baths, but the $w,c$ dissipators depend
in a {\it nonadditive} manner on the original 
coupling parameters, $\lambda_{w,c}$. 

We solve the Redfield equation in the energy basis
%in the steady-state limit, $\dot{\rho}_{ES}(t) = 0$ 
under the constraint of $\Tr[\rho_{ES}(t)] = 1$ for all $t$ 
and obtain the steady state density matrix of the extended system, $\rho_{ES}^{ss}$. 
The steady state heat current at the $\alpha$ contact is 
%\bea
$j_{\alpha}={\rm Tr}\left[ {\cal D}_{\alpha}(\rho_{ES}^{ss}) \hat H_{ES}^{D}\right]$;
%\eea 
%
the heat current is defined positive when flowing from the bath towards the system.
%This procedure, the application of the Redfield  QME onto the RC model
%is referred to as the reaction-coordinate quantum master equation method, RC-QME. 

%--------------------------------------------
% Figure 2
\begin{figure*}[hbt]
 \center
 \includegraphics[width=2\columnwidth]{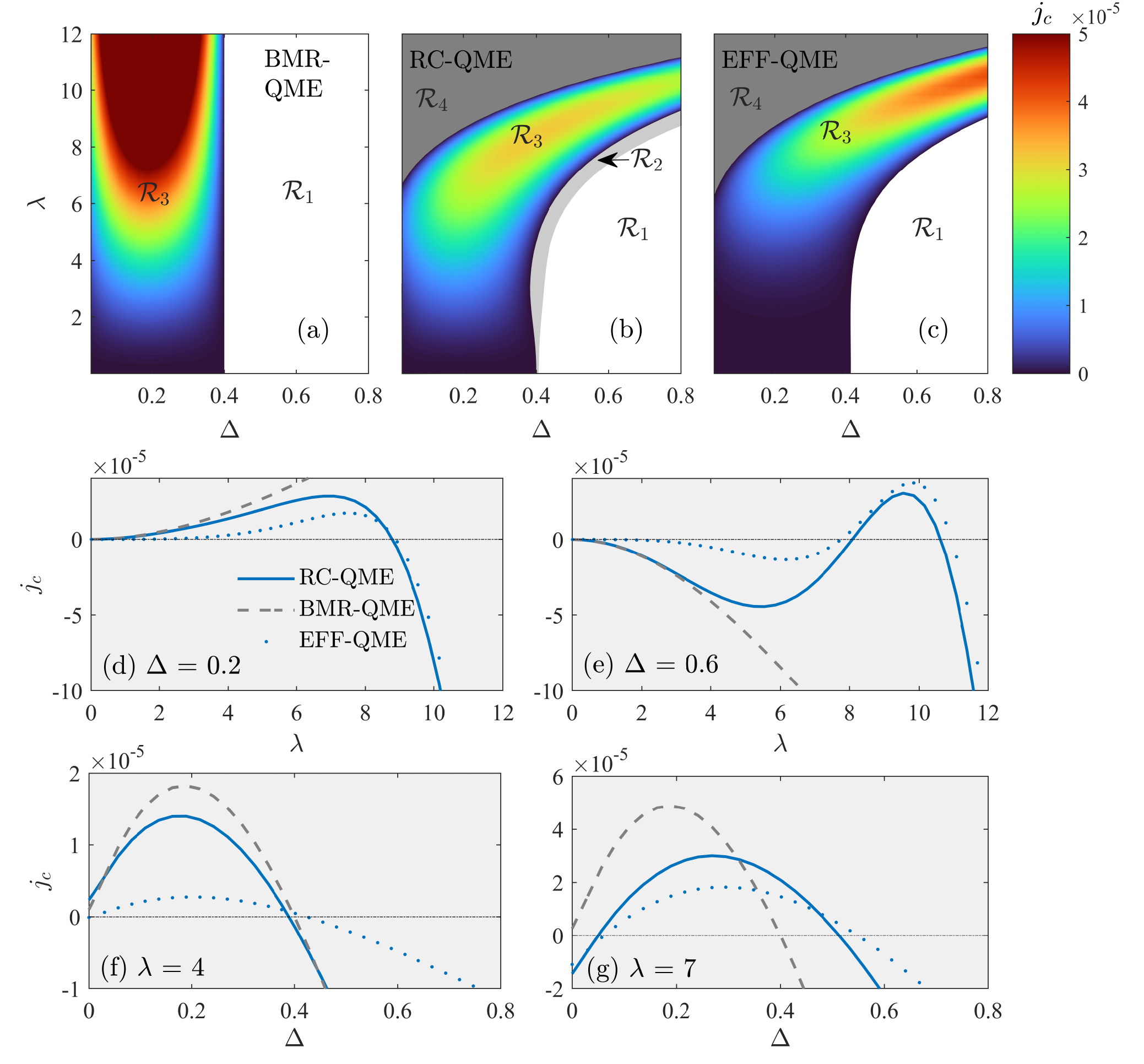}
\caption{(a)-(c) 
Contour maps of the cooling current $j_c$ as a function of the coupling strength $\lambda$ 
and energy spacing $\epsilon_2-\epsilon_1=\Delta$ using the
(a) BMR-QME,  (b) RC-QME with $M=6$,  (c) EFF-QME method. 
The system cools ($j_c>0$) in region $\mathcal R_3$ only. 
We further color $\mathcal R_2$ and $\mathcal R_4$ in which $j_w>0$ but $j_c<0$;
in $\mathcal R_1$ both $j_w<0$ and $j_c<0$.
(d)-(g) Cuts of contour maps, with $j_c$ as a function of (d) $\lambda$ with $\Delta = 0.2$, 
(e) $\lambda$ with $\Delta = 0.6$, 
(f) $\Delta$ with $\lambda = 4$, and 
(g) $\Delta$ with $\lambda = 7$
We  use BMR-QME (\dablack), RC-QME with $M=6$ (\sblack) and the EFF-QME (\dblack). 
Other parameters are   $\Omega=20$, $T_c = 0.25$, $T_h=0.5$, $T_w=1.5$, 
$\gamma = 0.0071/\pi$, $\Lambda_c = \Lambda_h = \Lambda_w = 500$.}
\label{fig:Fig2}
\end{figure*}
%NAS- Color bar, make it inline with the color plot. (put j_c elsewhere.)  - It wouldn't hurt to have the legend text size a little bit large in panels (d) and (e). - Maybe it is just too late when I'm writing this, but the colors white and yellow seem aggressive to the eye (at least for me), what about instead of white, try a shade of orange, and instead of yellow try maybe a dark green? In panel (e), the line for zero shouldn't be red if it is already used for \lambda = 2.
%NAS - I think we are best to split up the cuts at different parameter values. have \Delta = 0.2 be one plot, and another for \Delta = 0.6 (same for panel (e)). A main message of this work is the distinction between weak coupling and strong coupling techniques, as well as an effective model. So, having all these lines on the same plot might distract from the simple message. 
%NAS - Check the code, what exactly is g?
%NAS - Panel (e) why does it not start from zero?  

%------------------------------------------------------------------
%\subsection*{Effective model at large $\Omega$}
%\label{Sec: EffModel}
{\it Effective QAR model.--}
Assuming that the RC frequencies are high,
$\Omega \gg\Delta,\lambda,T$
%, yet $T< \Omega$, 
we can recreate a 3-level model for the refrigerator by truncating
the extended $3M^2$ (QAR with two RCs) system. 
This effective (EFF) model captures strong-coupling effects through level renormalization, a principle introduced in Ref. \cite{Nick_2021} for the nonequilibrium spin-boson model. 
%prominent aspects of the QAR-RC model can be captured 
The lowest three energy levels of the manifold of $\hat H_{ES}^D$
roughly correspond to the system occupying one of the levels $|1\rangle$, $|2\rangle$, 
or $|3\rangle$ [see Eq. (\ref{eq:RC-QAR Hamiltonian})], while the two RCs are in their respective ground state 
(energy $2\times \frac{1}{2}\Omega$); the next manifold, which is further away in energy,
involves a single RC excitation 
(energy $\frac{1}{2}\Omega +\frac{3}{2}\Omega \gg \Delta,\lambda$).
The effective model includes three levels, similarly to the original QAR Hamiltonian
before the incorporation of the RCs, Eq. (\ref{eq:QARHamiltonian}).
However, it includes
strong coupling effects through the $\lambda$-dependent eigenenergies 
and system-bath coupling matrix elements, see \cite{SuppI}.
We study steady state transport in this model
using the Redfield equation, and refer to these calculations as the EFF-QME approach. 
 
%In that work we investigated strong coupling effects in quantum thermal transport. We showed that the crossover behavior of the current with system-bath coupling could be  rationalized with the construction of an effective model.
%This model was constructed using the thermal inaccessibility of excited states of the RC harmonic oscillators. As such, this effective model constrains the system to occupy only its low lying states, resulting in an effective Hamiltonian with similar form to the original system, but with coupling strength dependent parameters.
%In this subsection, we motivate a similar procedure to produce an effective quantum absorption refrigerator (EFF-QAR) to rationalize strong coupling features of this device. 
%
%Following a procedure analogous to Ref. \cite{Nick2021}, we first obtain the spectrum of the extended RC-QAR system. The first 6 energy eigenvalues of said spectrum are shown in Figure \ref{fig: Eigenspectrum}.  % M=?
%The large separation of order $\Omega$ between the low lying and excited states is associated with having different RC occupation numbers. namely, the low lying states correspond to the RC occupying its ground state, while the 3 excited states correspond to one of the RCs being in its first excited state.
%Therefore, we treat these three lower levels as the EFF-QAR defining the coupling depedent energy level parameters as $\Delta_{eff}(\Vec{\lambda}) = E_2(\Vec{\lambda}) - E_1(\Vec{\lambda})$ and  $g{eff}(\Vec{\lambda}) = E_3(\Vec{\lambda}) - E_2(\Vec{\lambda})$. 

{\it Results.--}
%Before presenting numerical simulations, we as
What are the expected signatures of strong coupling on the performance of thermal machines?
We recount:
(i) {\it Renormalization of energy parameters}, leading to
the reshaping of the performance window \cite{Nick_2021}.
(ii) {\it Bath-cooperative transitions} \cite{Nicolin_2011,Hava_2019}, which can lead
to {\it leakage}, that is heat current flowing directly from the work bath to the cold one.
In this situation,  $j_w>0$ while $j_c<0$.
(iii) {\it Off-resonant transitions}
that open additional transport pathways. 
In the present model, strong coupling opens additional cooling cycles
that involve excitations of the RCs;
in the weak coupling limit, only a single cycle contributes, see Fig. \ref{fig:Fig1}.
While running a machine with multiple cycles may enhance the cooling window,
it results in  {\it internal dissipation}, with the system unable to reach the maximal Carnot efficiency 
\cite{Correa_2015,Hava_2019}. % DDD

To discern the relative importance of these factors,
we perform heat transport simulations of the strongly-coupled 3-level QAR using 
(i) The Born-Markov Redfield equation {\it after} performing the 
reaction coordinate transformation on the cold and work baths (RC-QME). 
Based on benchmarking  on the spin-boson model \cite{Nick_2021}, 
we expect this method to accurately describe strong coupling effects even at large $\lambda$,
as long as $\gamma$ is small and $\Omega$ is large. % DDDN 
% (facilitating convergence).
%This method shoudl capture the mechanism S1 to S3.
(ii) The standard Born-Markov Redfield equation without performing the RC transformation (BMR-QME).
This approach obviously fails to describe the correct behavior as one increases the coupling $\lambda$.
(iii) The Born-Markov Redfield equation on the {\it effective} QAR model 
(EFF-QME). This method presents strong $\lambda$ effects within the renormalization of energy parameters, yet missing leakage and  competing cycles.

We examine the cooling current and the window of operation in Fig. \ref{fig:Fig2}, presented 
as a function of the energy splitting $\Delta$ (see Fig. \ref{fig:Fig1}) and the
system-bath coupling  $\lambda$. 
The colored region ${\mathcal R_3}$ in the contour plots Fig. \ref{fig:Fig2} (a)-(c)
corresponds to the cooling window, $j_c>0$. 
In contrast, $j_c<0$ in regions ${\mathcal R_1}$, ${\mathcal R_2}$ and ${\mathcal R_4}$ 
(we explain their characteristics below).
Focusing on ${\mathcal R_3}$ and comparing weak coupling, Fig. \ref{fig:Fig2}(a), 
to RC-QME calculations, Fig. \ref{fig:Fig2}(b), we
arrive at the main conclusion of our work:
Reshaping of the cooling window is a central impact of strong coupling, 
and it results from parameter renormalization at strong coupling; we show energy shift and modulation
of couplings with $\lambda$ in Ref. \cite{SuppI}.
This observation is corroborated by noting that EFF-QME simulations in Fig. \ref{fig:Fig2}(c), 
which do not incorporate leakage and off-resonant effects, very well emulate strong coupling characteristics.

In the weak coupling limit, Fig. \ref{fig:Fig2}(a), the cooling window is defined by the condition
$\frac{\epsilon_2-\epsilon_1}{\epsilon_3 - \epsilon_1}\leq \frac{\beta_h-\beta_w}{\beta_c-\beta_w}$
 \cite{KosloffQAR}, which translates to $\Delta \leq 0.4$ in our parameters, independent of $\lambda$.
In the EFF-QAR model, the relevant measure for the cooling window is
the ratio $ \frac{E_2(\lambda)-E_1(\lambda)}{E_3(\lambda) - E_1(\lambda)}\leq
\frac{\beta_h-\beta_w}{\beta_c-\beta_w}$,  with $E_i(\lambda)$ as the first three eigenvalues of the 
extended system. % (where we highlight dependence on the system-bath coupling).
Indeed, this inequality well predicts the cooling window beyond weak coupling \cite{SuppI}.
%particularly, entering and leaving the cooling window as a function of $\lambda$ e.g. for $\Delta =0.6$.
%If we begin form  $\Delta =0.2$, which is inside the cooling window at weak coupling,
%we find that $0\leq R_{\lambda}\leq 0.4$  
%so long as $\lambda\lesssim 7$  \cite{SuppI}.
%This indicated that the cooling window should be terminated at that value, 
%as we indeed observe in  Fig. \ref{fig:Fig2}(b).
%Similarly, if we begin outsude the weak-coupling cooling window at $\Delta =0.6$ 
%we find that $R_{\lambda}<0.4$ once $4\leq\lambda\leq 8$,
%thus the system enters, then leaves, the cooling regime at intermediate couplings,
%as indeed we observ in Fig. \ref{fig:Fig2}(b).
%
The remarkable success of the EFF-QAR model (compare Fig. \ref{fig:Fig2}(c) to  \ref{fig:Fig2}(b))
evinces that renormalization of parameters is a central fingerprint of strong system-bath coupling.
The impact of heat leakage ($j_c<0$ but $j_w>0$) is further identified
in Fig. \ref{fig:Fig2}(b) by ${\mathcal R_2}$ and ${\mathcal R_4}$:
While within ${\mathcal R_2}$, ordering of levels still follows the $\lambda=0$ limit,
in region ${\mathcal R_4}$, parameter-shift is significant such that 
the level corresponding to  $\epsilon_2$ crosses the one originating from $\epsilon_1$, see \cite{SuppI}.
Finally, in region ${\mathcal R_1}$ both $j_w<0$ and $j_c<0$, with heat arriving from the hot bath.

\begin{figure}[h!]
\centering
\includegraphics[width=\columnwidth]{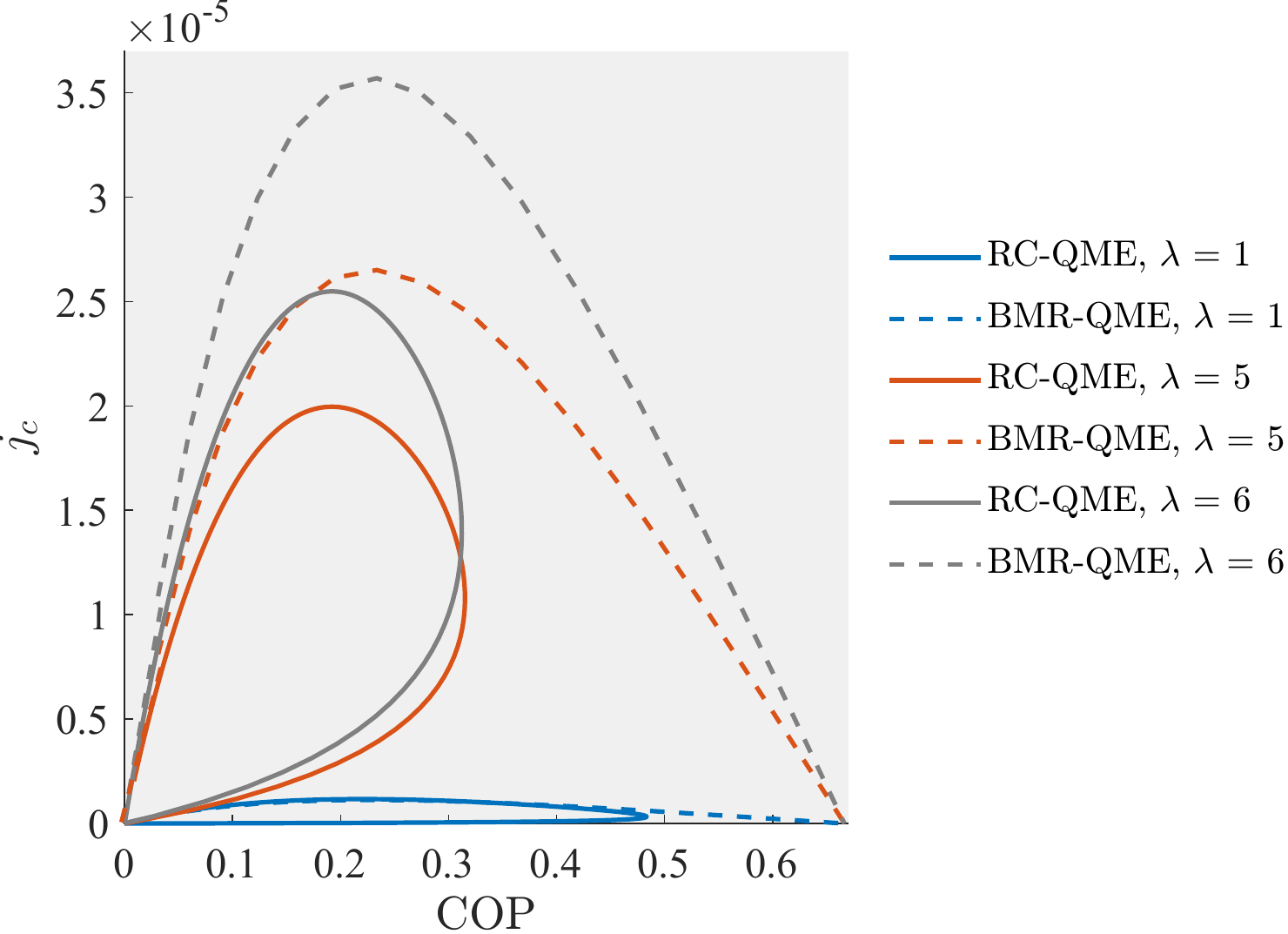}
\caption{Cooling current ($j_c$) as a function of the coefficient of performance (COP) with
the BMR-QME (dashed) and RC-QME (full) approaches at different values for $\lambda$.
%The blue dashed line indicates the the upper bound on the COP predicted from weak coupling thermodynamics. 
    %Here dotted (\dblack), dashed (\dablack),  and dash-dotted (\ddotblack) lines indicate RC-QME results with $\lambda = 1, 2,$ and $5$ respectively. 
Other parameters are the same as in Fig. \ref{fig:Fig2}. }
\label{fig: Fig4}
\end{figure}

The  relationship between the cooling current, $j_c$, and the coefficient of performance (COP), 
$j_c/j_w$, is displayed in Fig. \ref{fig: Fig4}.
In the weak coupling limit, the 3-level model can reach Carnot's bound 
$\eta_C=(\beta_h-\beta_w)/(\beta_c-\beta_w)$, 
albeit at a vanishing cooling power \cite{KosloffQAR}. 
This idealized scenario is applicable only for reversible thermal machines, in the absence of 
heat leaks and internal dissipation.
In contrast, at stronger coupling these processes become active,
thus suppressing the COP.
Moreover, at strong coupling the cooling current is smaller than the weak coupling value,
demonstrating the negative impact of heat leaks (direct heat flow
from the work to the cold bath).
%Instead, such endeavours should attempt to construct an optimal QAR inside the cooling window where coupling to the environment is weak. This concludes our discussion.

%============================================================
% Figure 4

%=====================================================================================
{\it Summary.}
We revealed that a prominent signature of strong system-bath coupling in quantum thermal machines
is the modification, or reshaping of the operational window, 
arising due to parameter renormalization.
Focusing on the 3-level QAR, we reached this conclusion based on comprehensive simulations made 
possible with the non-perturbative RC-QME method.
This observation, on the central role of parameter renormalization, was corroborated by
the analysis of an {\it effective} 3-level QAR model, and it should hold for other types of spectral functions.
%Other effects such as heat leaks were further demonstrated to arise at strong couplings. 
%
%In this model,
%parameters were renormalized due to strong couplings, yet
%the tight coupling limit was obeyed for any $\lambda$, $j_c\propto j_w$.
%To a large extent, the effect of strong system-bath coupling is to reshape the cooling window.
Detrimental effects of strong couplings are due to heat leaks and internal dissipation, 
which reduce the cooling current and the COP.
The design of quantum thermal machines that markedly benefit from strong system-bath coupling remains an
open challenge.

%By tuning the coupling strength, this effect shows benefit by opening a region of cooling otherwise prohibited in the BMR limit. The second mode of transport, generally detrimental to cooling due to its entropic nature, is activated by strong coupling via direct inter-bath interaction. In sum, strong coupling in QARs is not economically desirable, drawing on its inverse relationship with COP.

% sequential/tigh couoking transport in the effective model. 
% inter-bath transitno

%================================

%================================
\begin{acknowledgements}
DS acknowledges support from an NSERC Discovery Grant and the Canada Research Chair program.
The work of FI was supported by a CQIQC summer fellowship at the University of Toronto.
\end{acknowledgements}

%=================================================================
%\bibliographystyle{unsrt.bst} 
%\section*{References}
%\bibliographystyle{iopart-num}
%\bibliography{bibliographyQAR}
%=======================================

%===============
\end{document}

\title{Supplemental Material: Strong system-bath coupling reshapes characteristics of quantum thermal machines}

\author{Felix Ivander}
\altaffiliation{These authors contributed equally to this work}
\affiliation{Chemical Physics Theory Group, Department of Chemistry and Centre for Quantum Information and Quantum Control,
University of Toronto, 80 Saint George St., Toronto, Ontario, M5S 3H6, Canada}
\author{Nicholas Anto-Sztrikacs}
\altaffiliation{These authors contributed equally to this work}
\affiliation{Department of Physics, 60 Saint George St., University of Toronto, Toronto, Ontario, Canada M5S 1A7}
\author{Dvira Segal}
\affiliation{Chemical Physics Theory Group, Department of Chemistry and Centre for Quantum Information and Quantum Control,
University of Toronto, 80 Saint George St., Toronto, Ontario, M5S 3H6, Canada}
\affiliation{Department of Physics, 60 Saint George St., University of Toronto, Toronto, Ontario, Canada M5S 1A7}
\email{dvira.segal@utoronto.ca}

%==============================================
\date{\today}
\maketitle

\begin{widetext}
\vspace{5mm}
\renewcommand{\theequation}{S\arabic{equation}}
\renewcommand{\thefigure}{S\arabic{figure}}
\renewcommand{\thesection}{S\arabic{section}}
\setcounter{equation}{0}  % reset counter
\setcounter{figure}{0}

%=================================================================
In this Supplemental Material, we provide additional details on the application of the
reaction-coordinate (RC) mapping on the 3-level quantum absorption refrigerator (QAR) model and 
the computational methods used. 
Sec. \ref{S0} provides a top-level picture of the methods adopted in the main text.
In Sec. \ref{S1}, we detail steps in arriving at the
reaction coordinate quantum master equation  (RC-QME) framework.
%we clarify and expand upon strong coupling effects arising in the quantum absorption refrigerator (QAR). % DDD sentence not informative 
%Namely, in Sec. \ref{S1} we provide a thorough discussion surrounding the application of the Born-Markov Redfield quantum master equation in the reaction coordinate framework (RC-QME). 
% DDD thorough discussion --> make it more concrete
%
In Sec. \ref{S2}, we justify the construction of the effective 3-level QAR model,
which incorporates strong couplings with renormalized parameters only.
%Sec. \ref{S3} provides a brief summery of the methods we use.
We discuss aspects of convergence of the RC-QME method in Sec. \ref{S4}.
%In Sec. \ref{S2}  we give additional background surrounding the effective quantum absorption refrigerator (EFF-QAR) model. % DDD make it concrete!
 %Furthermore, we showcase results of the truncation scheme used to represent the reaction coordinate harmonic oscillators as $M$-level systems in Sec. \ref{S3}.  % DDD too detailed for this high level sentence
 %
In Sec. \ref{S5}, we present additional simulations of the 3-level QAR 
in different parameters regimes, discussing 
the cooling current and the coefficient of performance (COP).
 %an additional heat flow mechanism that is available at strong coupling.  % DDD too detailed for this high level paragraph.

%===============================================
\section{Diagram of methods}
\label{S0}

%==============================================================
% Figure S0
\begin{figure}[h]
\centering
\includegraphics[width=0.5\columnwidth]{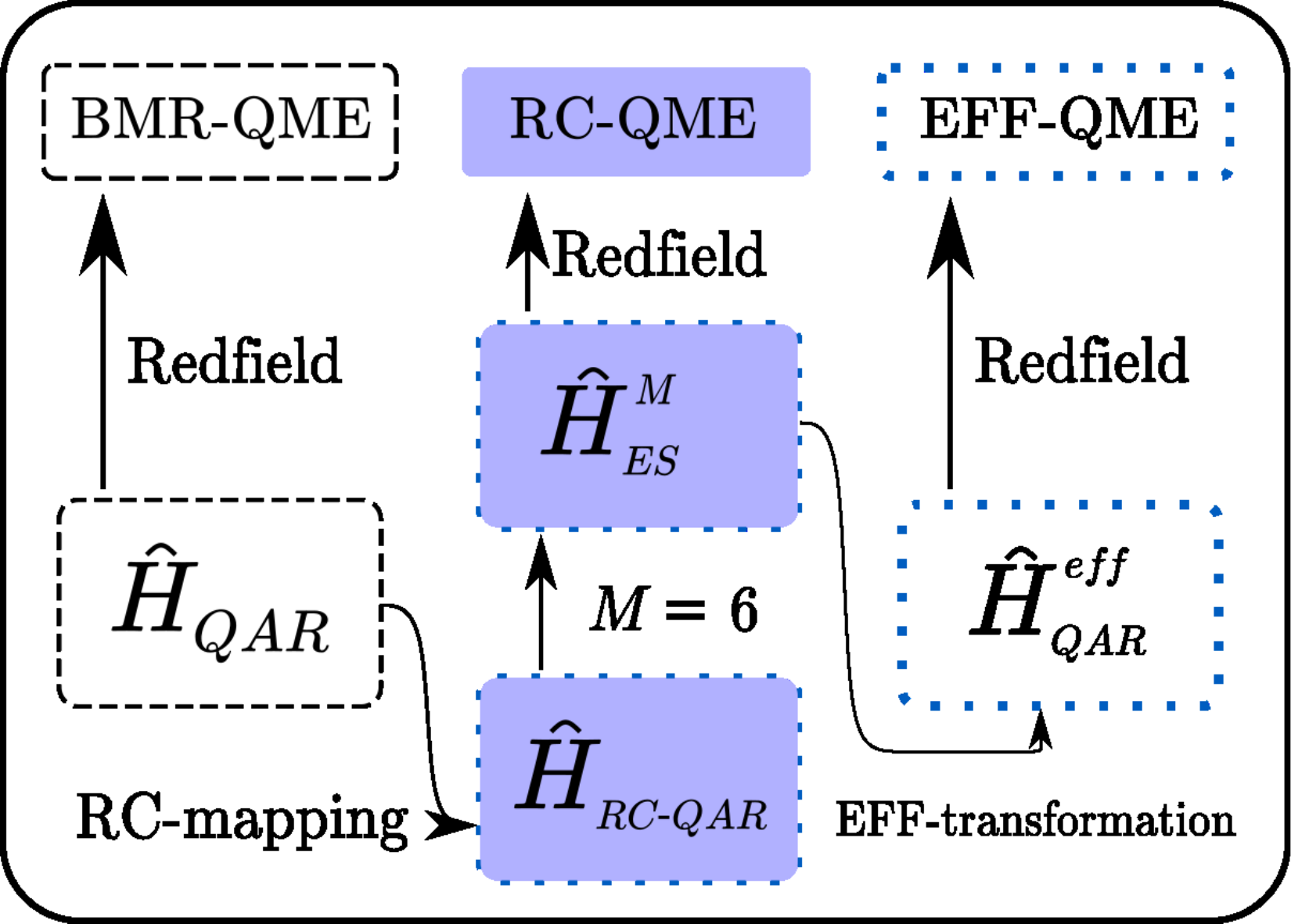}
\caption{
A schematic flowchart of methods used in this study.
Objects enclosed in dashed lines (left side) concern the BMR-QME technique,
where the Redfield equation is used under the weak coupling approximation.
Objects in shaded blue (center) show the development of the RC-QME method (e.g., with $M=6$),
a strong coupling treatment.
Objects within dotted boxes (right) lead to the EFF-QME method, a truncated form of the RC-QME method.}
%effective QAR with coupling strength dependent parameters where the standard Redfield equation can be used and strong coupling effects will emerge.}
\label{fig:diagram}
\end{figure}
%================================
In Fig. \ref{fig:diagram}, 
we summarize the three approaches taken in this work:

(i) The RC-QME method allows us to uncover strong coupling effects in the performance of QARs.
We start with the Hamiltonian $\hat H_{QAR}$. We use the RC mapping and incorporate two 
RCs into the system, thus reaching $\hat H_{RC-QAR}$. We then
truncate the resulting extended model to e.g. include $M=6$ levels for each RC, arriving at
$\hat H_{ES}^M$. We study the open dynamics of this model 
(in the eigenenergy representation of the extended system) using the Redfield QME.

(ii) The effective quantum master equation (EFF-QME) method approximates the RC-QME method.
Here, we build an effective 3-level model from $\hat{H}_{ES}^M$ by truncating the manifold, then proceed with the Redfield QME.

(iii)
We further apply the Born-Markov-Redfield  quantum master equation (BMR-QME) directly on $\hat H_{QAR}$;
this method is valid only at weak coupling.

\vspace{4mm}

%==========================================================================================

\section{The RC-QME framework}
\label{S1}
We describe here the RC-QME procedure, a method non-perturbative in the system-bath coupling $\lambda$.
The 3-level QAR Hamiltonian is manipulated in several steps
before applying the Born-Markov Redfield quantum master equation.
This includes the extraction of the RC coordinates, truncation of the spectrum of the RC modes, 
and diagonalization of the resulting extended system Hamiltonian.
The starting point (see main text) is the 3-level QAR model given by the Hamiltonian 
\bea
\label{eq: QAR Hamiltonian}
\hat H_{QAR} = \sum_{i = 1}^3 \epsilon_i \ket{i}\bra{i} 
 + \sum_{\alpha=\{c,w,h\}}\sum_k \nu_{\alpha,k} \left( \hat c_{\alpha,k}^{\dagger} + 
\hat S_{\alpha}\frac{f_{\alpha,k}}{\nu_{\alpha,k}} \right)\left( \hat c_{\alpha,k} + \hat S_{\alpha}\frac{f_{\alpha,k}}{\nu_{\alpha,k}} \right).
\eea
Here, $\epsilon_i$ denotes the $i$th energy level of the QAR. Without loss of generality, we set the total energy gap to be $\epsilon_3 - \epsilon_1 = 1$ and the first energy gap as $\epsilon_2 - \epsilon_1 = \Delta$. The system operators $\hat S_\alpha$ with $\alpha=c,h,w$ describe the coupling of the QAR to 
the $\alpha$th bath, 
\bea
    \hat S_c = \ket{1}\bra{2} + h.c, \,\,\,\
    \hat S_h = \ket{1}\bra{3} + h.c, \,\,\,
    \hat S_w = \ket{2}\bra{3} + h.c. 
\eea
The baths comprise collection of harmonic modes with
creation (annihilation) operators, $\hat c_{\alpha,k}^{\dagger}$ ($\hat c_{\alpha,k}$), and
the system-bath couplings are captured by the spectral density function, 
$J_{\alpha}(\omega) = \sum_{k}f_{\alpha,k}^2\delta(\omega-\nu_{\alpha,k})$, with real-valued $f_{\alpha,k}$  the coupling energy between the QAR and a bath mode of frequency $\nu_{\alpha,k}$. 

\subsection{RC mapping}
To handle strong coupling effects between the system (3-level model) 
and the cold and work baths, we invoke the reaction coordinate mapping for those reservoirs. 
This results in the following RC-QAR Hamiltonian 
\bea
\hat H_{RC-QAR} &=&  \sum_{i = 1}^3 \epsilon_i \ket{i}\bra{i} +
\sum_{\alpha=\{c,w\}} \Omega_{\alpha} \hat a_{\alpha}^\dagger \hat a_{\alpha} 
+\sum_{\alpha=\{c,w\}} \hat S_{\alpha} \lambda_{\alpha}(\hat a_{\alpha}^\dagger+\hat a_{\alpha}) 
+ \sum_{\alpha=\{c,w\}} \frac{g_{\alpha,k}^2}{\omega_{\alpha,k}} (\hat a_{\alpha}^{\dagger}+\hat a_{\alpha})^2
\nonumber \\
&+& \sum_{k} \nu_{h,k} \left( \hat c_{h,k}^{\dagger} + \hat S_h\frac{f_{h,k}}{\nu_{h,k}} \right)
\left( \hat c_{h,k} + \hat S_h\frac{f_{h,k}}{\nu_{h,k}} \right)
+ \sum_{\alpha=\{c,w\}} (\hat a_{\alpha}^{\dagger}+\hat a_{\alpha})\sum_k g_{\alpha,k}
 (\hat b_{\alpha,k}^{\dagger}+\hat b_{\alpha,k}) 
\nonumber \\ 
&+& \sum_{k} \nu_{h,k} \hat c_{h,k}^\dagger \hat c_{h,k}
+ \sum_{\alpha=\{c,w\};k} \omega_{\alpha,k} \hat b_{\alpha,k}^\dagger \hat b_{\alpha,k}.
    \label{eq:RC-QAR Hamiltonian }
\eea
This mapping was reviewed in Ref. \cite{Nazir18}. It was described in details for the related spin-boson Hamiltonian in 
Refs. \cite{NazirPRA14,Nick_2021}, and we do not repeat those steps here.
 
As outlined in the main text, 
this model now describes a QAR coupled via $\lambda_{c,w}$ to two reaction coordinates of frequencies $\Omega_{c,w}$ denoted by the bosonic creation (annihilation) operators $\hat a_{c,w}^{\dagger}$ ($\hat a_{c,w}$) extracted from the cold and work baths respectively. These reaction coordinates couple to residual cold and work baths described by the operators $\hat b_{\alpha,k}^{\dagger}$ ($\hat b_{\alpha,k}$). The Hamiltonian Eq. (\ref{eq:RC-QAR Hamiltonian }) also contains the unaltered hot bath and its coupling to the QAR. The last term in the first line of Eq. (\ref{eq:RC-QAR Hamiltonian })  is quadratic in the RC operators. In what follows, we neglect this term as it was shown to be partially eliminated from the Markovian quantum master equation \cite{NazirPRA14,Nick_2021}. 
%For convenience, we assume a symmetry between the baths such that $\Omega_c = \Omega_w = \Omega$, $\lambda_c = \lambda_w = \lambda$ and $\gamma_{c,h,w} = \gamma$. 
In the RC representation, the spectral density functions describing the 
interaction between the baths and the system are Ohmic, 
$J_{\alpha}(\omega)=\gamma_{\alpha}\omega e^{-|\omega|/\Lambda_{\alpha}}$ for all baths.
%\eea
%
Here, $\Lambda_{\alpha}$ is the high-frequency cutoff of the $\alpha$th bath.
$\gamma_{\alpha}$ are dimensionless coupling constants between the extended system (ES), which comprises the original three levels and the two RCs, and the baths. In what follows, these constants are assumed small to justify employing the Redfield quantum master equation technique on the extended (three levels and two RCs) system. 

\subsection{RC truncation}
In order to employ the Redfield quantum master equation on the extended system, 
we truncate the RC manifolds, reducing them to their first $M$ levels. 
This results in a $3\times M^2$ dimension Hamiltonian of the extended system,  $\hat{H}_{ES}^M$.
The total Hamiltonian includes the truncated system along with the three thermal baths 
and the interactions between them,
\bea
\hat{H}_{RC-QAR}^M = \hat{H}_{ES}^M + \hat{H}_B + \hat{H}_{ES,B}^M.
\eea
Explicitly, 
\bea
\label{eq: HES}
\hat{H}_{ES}^M = \sum_{i=1}^3 \epsilon_i \ket{i}\bra{i} + \sum_{\alpha = \{c,w\}}\sum_{l_{\alpha}=0}^{M-1} \Omega_{\alpha}\left( l_{\alpha} + \frac{1}{2} \right) \ket{l_{\alpha}}\bra{l_{\alpha}} + \sum_{\alpha= \{c,w\}} \lambda_{\alpha} \hat{S}_{\alpha} \sum_{l_{\alpha}=1}^{M-1} \sqrt{l_{\alpha}} \ket{l_{\alpha}}\bra{l_{\alpha}-1} + h.c.
\eea
with $|l_{\alpha}\rangle$ the eigenstates of the $\alpha$ harmonic RC mode.
The baths Hamiltonian is given by

\bea
\hat{H}_B = \sum_{\alpha = \{c,w\}}\sum_k \omega_{\alpha,k} \hat{b}^{\dagger}_{\alpha,k} \hat{b}_{\alpha,k} + \sum_k \nu_{h,k} \hat{c}^{\dagger}_{h,k} \hat{c}_{h,k}, 
\eea
and the interaction term can be formally represented as
\bea
%\hat{H}_{ES,B}^M = \sum_{\alpha} \hat{V}_{ES,\alpha}^M \otimes \hat{B}_{\alpha} + \hat{S}_h^M\otimes\hat{B}_h.
\hat{H}_{ES,B}^M = \sum_{\alpha=\{c,w,h\}} \hat{V}_{ES,\alpha}^M \otimes \hat{B}_{\alpha}.
\eea
In this expression, $\alpha$ goes over the three heat baths---albeit with the hot bath showing a 
coupling form distinct from the cold and work baths, since it did not undergo the RC mapping,
\bea
\hat{V}_{ES,\alpha}^{M} &=& \sum_{l_{\alpha}=1}^{M-1} \sqrt{l_{\alpha}} (\ket{l_{\alpha}}\bra{l_{\alpha}-1} + \ket{l_{\alpha}-1}\bra{l_{\alpha}}),   \,\,\,\,\,
\hat{B}_{\alpha} = \sum_k g_{\alpha,k} (\hat{b}^{\dagger}_{\alpha,k} + \hat{b}_{\alpha,k}),
\,\,\,\ {\rm for }\,\,\alpha=w,c
\nonumber\\
\hat{V}_{ES,h}^M &=&\hat{S}_h, % \otimes1_c^M\otimes1_w^M, % DDD inconsistent with the rest of the text
%\\
%\hat{B}_{\alpha} &&= \sum_k g_{\alpha,k} (\hat{b}^{\dagger}_{\alpha,k} + \hat{b}_{\alpha,k}),
%\\
\,\,\,\, \hat{B}_{h} = \sum_k f_{h,k}(\hat{c}^{\dagger}_{h,k} + \hat{c}_{h,k}).
\eea
%
%Here, $1_\alpha^M$ is the $M \times M$ identity operator of the $\alpha$th reaction coordinate.

\subsection{Diagonalization of the extended system}
In order to study dynamics and transport with the
 Redfield quantum master equation, we diagonalize the truncated-extended system Hamiltonian, Eq. (\ref{eq: HES}). 
This step is typically done numerically. The diagonalization involves the unitary operator $\hat U$,
\bea
\hat{H}_{ES}^D = \hat U^{\dagger}\hat{H}_{ES}^M\hat U, 
\hspace{1cm} \hat{V}_{ES,\alpha}^D = \hat U^{\dagger}\hat{V}_{ES,\alpha}^M\hat U.
\label{Eq: Diagonalized Hamiltonian}
\eea
We denote the eigenenergies of $\hat{H}_{ES}^M$ by $E_{n}(\vec{\lambda})$; $n=1,2...,3M^2$, 
and present the first six of them in Fig. \ref{fig:S1}. %%%DDD
Additionally, in Fig. \ref{fig:S2} we display the system-bath coupling in the energy eigenbasis.
As an example, we display several matrix elements,
$|\langle m(\vec{\lambda}) | V_{ES,\alpha=w}^D |n(\vec{\lambda})\rangle|$ with 
$|n (\vec{\lambda})\rangle$ and $|m(\vec{\lambda})\rangle$
eigenvectors of $\hat{H}_{ES}^M$; we group the coupling parameters $\lambda_\alpha$ with a vector notation
$\vec{\lambda} = (\lambda_c,\lambda_w)$.

%==================================
% Figure S1
\begin{figure}[ht!]
\includegraphics[width=0.5\linewidth]{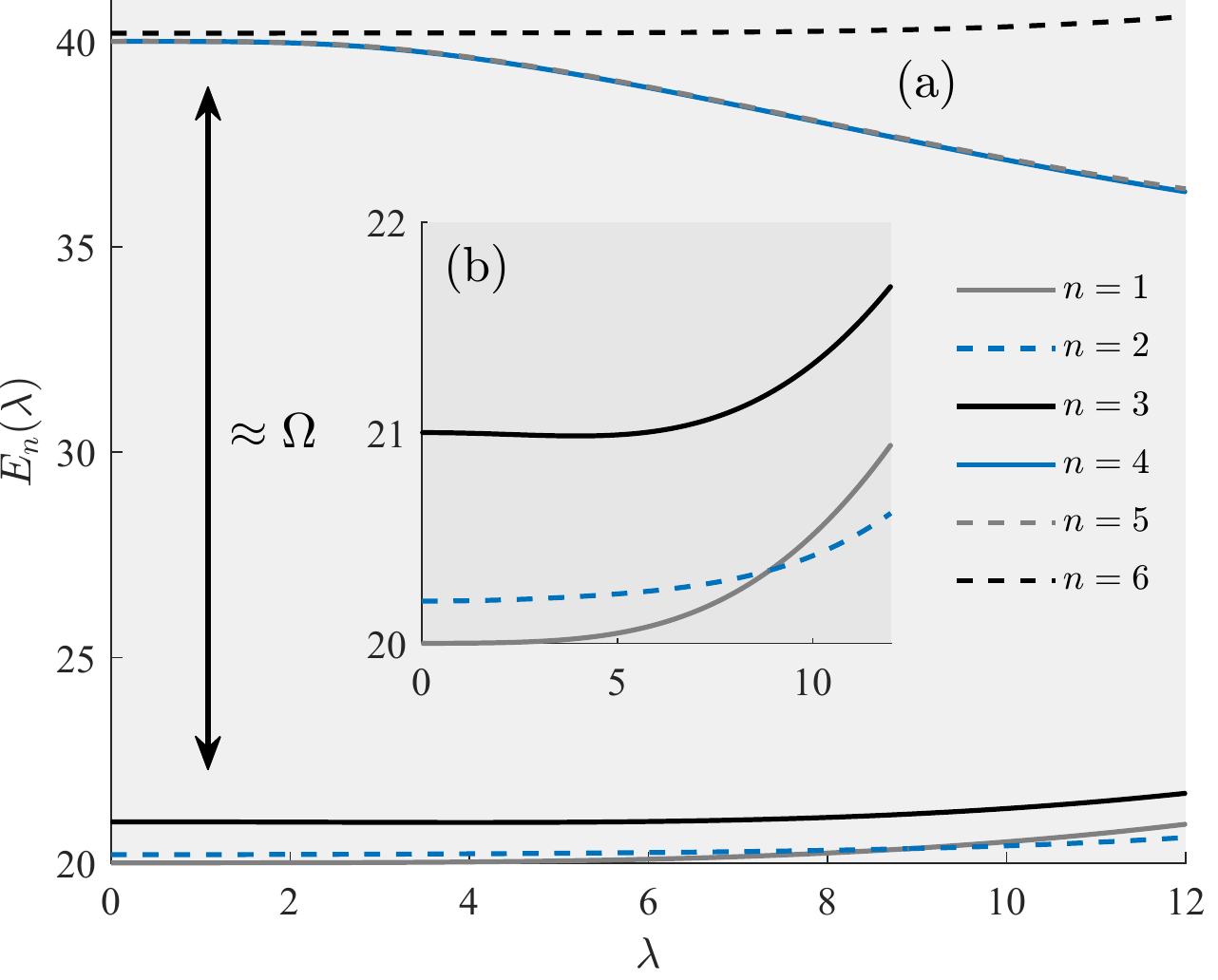}
\caption{
(a) The first 6 eigenenergies of the extended system Hamiltonian $\hat{H}_{ES}^M$, 
Eq. (\ref{eq: HES}). 
(b) The first 3 eigenenergies zoomed for clarity. 
Parameters are $\epsilon_1=0$; $\epsilon_2=\Delta= 0.2$, $\epsilon_3=1$
and RCs with $\Omega = 20$ and $M = 6$ levels.} %$T_c = 0.25$, $T_h=0.5$, $T_w=1.5$.}
  \label{fig:S1}
\end{figure}

\begin{figure}[ht!]
\includegraphics[width=0.5\columnwidth]{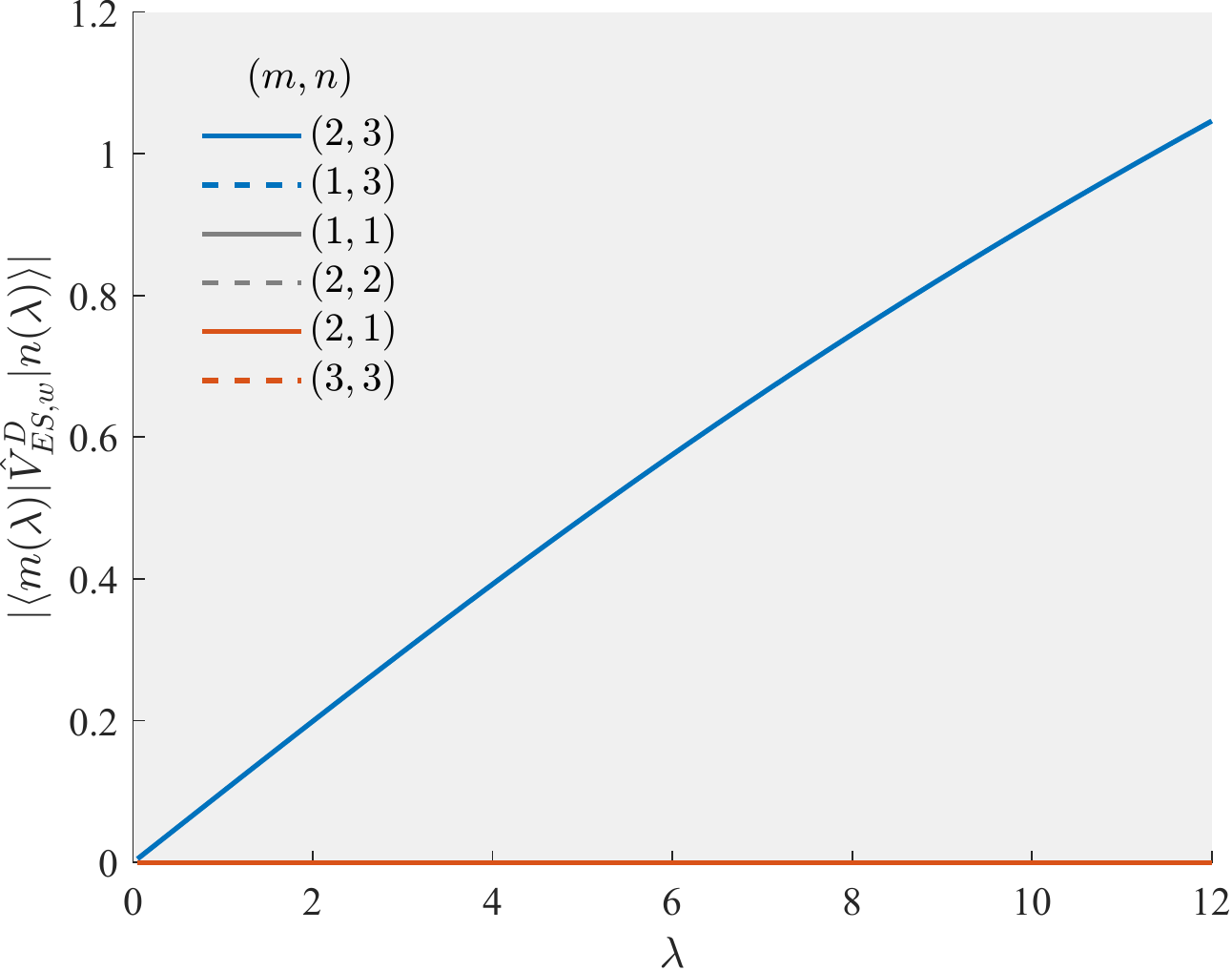}
\caption{Magnitudes of matrix elements of $\hat{V}_{ES,w}^D$ with $M=6$. 
Here $|m(\vec{\lambda})\rangle$ and $|n(\vec{\lambda})\rangle$ are eigenvectors of $\hat{H}_{ES}^M$.
In the present parameters---for very large $\Omega$--- 
the work bath dominantly couples to a single transition, between the second and third eigenstates,
while the other transitions are close to zero.
Parameters are identical to Fig. \ref{fig:S1}.}
 \label{fig:S2}
\end{figure}

%================================

\subsection{Redfield equation}

The Redfield quantum master equation 
for the dynamics of the extended system density matrix, $\rho_{ES}$, 
can be written in the Schr\"odinger picture as \cite{Nitzan}
\bea
\dot{\rho}_{ES,mn}(t) &=& -i\omega_{mn} \rho_{ES,mn}(t)
\nonumber\\ &+&
\sum_{\alpha}\sum_{j,k}[
R_{mj,jk}^{{\alpha}}(\omega_{kj}) \rho_{ES,kn}(t) + R_{nk,kj}^{{\alpha}*}(\omega_{jk}) \rho_{ES,mj}(t)
\nonumber\\ &-&
R_{kn,mj}^{{\alpha}}(\omega_{jm}) \rho_{ES,jk}(t) - R_{jn,mk}^{{\alpha}*}(\omega_{kn}) \rho_{ES,jk}(t)].
\label{eq: Redfield}
\eea
Here, $\omega_{mn}=E_m(\vec{\lambda})-E_n(\vec{\lambda})$ are the eigenenergies of the extended system.
The elements of the $R$ super-operator are given by half Fourier transform
of bath autocorrelation functions \cite{Nitzan},
\bea
R_{mn,jk}^{{\alpha}}(\omega) &=& (V_{ES,{\alpha}}^{D})_{m,n} (V_{ES,{\alpha}}^{D})_{j,k} \int_0^{\infty} d\tau e^{i\omega\tau} \langle B_{\alpha}(\tau)B_{\alpha}\rangle
\nonumber\\ &=&
(V_{ES,{\alpha}}^{D})_{m,n} (V_{ES,{\alpha}}^{D})_{j,k} [\Gamma_{{\alpha}}(\omega) + i\Delta_{{\alpha}}(\omega)].
\label{eq:dissipator}
\eea
These autocorrelation functions are evaluated with respect to the thermal state of their respective baths.
In what follows, we neglect the imaginary component of the dissipators, $\Delta_{\alpha}$, 
as it can be shown that they cancel out the quadratic term in the RC Hamiltonian, 
Eq. (\ref{eq:RC-QAR Hamiltonian }) \cite{NazirPRA14}.
For harmonic environments and a bilinear coupling between the extended system and the
residual environments, the real part of the $R$ tensor reduces to
\bea
\Gamma_{{\alpha}}(\omega) =
\begin{cases}
\pi J_{{\alpha}}(\omega) n_{\alpha}(\omega) & \omega > 0   \\
\pi J_{{\alpha}}(|\omega|)[(n_{\alpha}(|\omega|) + 1] & \omega < 0. \\
\pi\gamma_{\alpha}/ \beta_{\alpha}& \omega = 0
\end{cases}
\eea
$n_{\alpha}(\omega)$ is the Bose-Einstein distribution function of the ${\alpha}$th bath 
at inverse temperature $\beta_{\alpha}$;
$J_{{\alpha}}(\omega)$ is the spectral function of bath $\alpha$.
In the RC-QME method as described here, these functions are all ohmic.

To calculate heat currents flowing through the system, we write down the Redfield equation as
\bea
\dot \rho_{ES}(t)=-i[\hat H_{ES}^{D},\rho_{ES}] + {\mathcal D}_w(\rho_{ES}) + 
{\mathcal D}_h(\rho_{ES}) + {\mathcal D}_c(\rho_{ES}),
\eea
with the dissipators ${\mathcal D}_{w,h,c}(\rho_{ES})$ prepared based on Eq. (\ref{eq: Redfield}). 
These dissipators are additive given the weak coupling approximation 
of the RCs to their residual baths.
However, the dissipators ${\mathcal D}_{\alpha=w,c}$ 
depend in a non-additive way on the original coupling parameters, $\lambda_{w,c}$. % DDD
We solve this equation in steady state to obtain the long time solution, $\rho_{ES}^{ss}$.
The steady state heat current at the $\alpha$ contact is
%\bea
$j_{\alpha}={\rm Tr}\left[ {\cal D}_{\alpha}(\rho_{ES}^{ss}) \hat H_{ES}^{D}\right]$.
%\eea 
%

%======================================================================
\section{Effective Quantum Absorption Refrigerator model}
\label{S2}

%===============================================

A powerful aspect of the RC mapping as described in Sec.  \ref{S1}
is that it allows the construction of an effective low-dimension QAR model (EFF-QAR), which
takes into account strong coupling effects in the shift (renormalization) of parameters
relative to the weak-coupling limit.
This effective model is applicable when  $\Omega_{\alpha}$ is large
 ($\Omega_{\alpha} \gg \lambda_{\alpha},T_{\alpha},\Delta$). 
In this regime, dynamics is determined to a large extent by the 
lowest 3 eigenstates  of $\hat H_{ES}^M$
%the truncation of the extended system Hamiltonian reduces to a $3\times M^2$ systema, where the key features of the model are accessible through considerations of only the lowest  3 energy levels. In this regime, 
since excitations of the RC oscillators beyond their ground state are small.
The effective model is constructed as follows. Once
the diagonalization is performed in Eq. (\ref{Eq: Diagonalized Hamiltonian}) on a large enough $M$ (where convergence is achieved), 
we take into account only the three lowest eigenstates of the extended system (\ref{eq: HES})
%where the system parameters are renormalized due to the effects of strong coupling brought about by the reaction coordinates. %This approximate mapping is valid for $\Omega_{\alpha} \gg T_{\alpha}$ where the excited states of the reaction coordinates are thermally inaccessible.
and write down the effective QAR Hamiltonian as
\bea
\nonumber
\hat{H}_{QAR}^{eff} &=& \sum_{n=1}^3  E_n(\vec{\lambda}) \ket{n(\vec{\lambda})}\bra{n(\vec{\lambda})}  
+ \sum_{\alpha} F_{\alpha}(\vec{\lambda}) \hat{S}_{\alpha} \sum_k g_{\alpha,k} (\hat{b}^{\dagger}_{\alpha,k}+\hat{b}_{\alpha,k})+ \hat{S}_h\sum_k f_{h,k} (\hat{c}^{\dagger}_{h,k}+\hat{c}_{h,k}) 
\\
&+& \sum_{\alpha,k} \omega_{\alpha,k} \hat{b}^{\dagger}_{\alpha,k}\hat{b}_{\alpha,k} + \sum_{k} \nu_{h,k} \hat{c}^{\dagger}_{h,k}\hat{c}_{h,k}.
\label{eq:EFF-QAR}
\eea 
The energy levels of the QAR are $E_1(\vec{\lambda})$, $E_2(\vec{\lambda})$ and $E_3(\vec{\lambda})$,
which should be compared to $\epsilon_{1}$, $\epsilon_2$, $\epsilon_3$, respectively, the latter the levels
in the original QAR picture.
%Energy gaps between states of the QAR are related to the difference in eigenvalues of the extended system Hamiltonian Eq. (\ref{eq: HES}), with the first energy gap given by $E_2(\vec{\lambda}) - E_1(\vec{\lambda})$ and the second gap via $E_3(\vec{\lambda}) - E_1(\vec{\lambda})$. 
The eigenvectors associated with the energy levels $E_n(\vec{\lambda})$ 
are $\ket{n(\vec{\lambda})}$. %$n=\{,b,c\}$ labelling the eigenstates. 
The dimensionless real-valued 
factors $F_{\alpha}(\vec{\lambda})$  % DDD REAL? or do we need to seprate S to to the two components?
that decorate the system-bath coupling operators for
the work and cold baths
are  $F_c(\vec{\lambda}) = \bra{1(\vec{\lambda})}\hat{V}_{ES,c}^M\ket{2(\vec{\lambda})}$
and $F_w(\vec{\lambda}) = \bra{2(\vec{\lambda})}\hat{V}_{ES,w}^M\ket{3(\vec{\lambda})}$.
%
%These prefactors depict the dressing of the system operator with coupling strengths $\lambda$. 
We can then define a new spectral density function for the cold and work baths, 
scaled by $F_{\alpha}(\vec{\lambda})$,
\bea
J_{QAR,\alpha}^{eff}(\omega) = \gamma_{\alpha}\omega\left[F_{\alpha}(\vec{\lambda})\right]^2 e^{-|\omega|/\Lambda_{\alpha}} 
%= \left[F_{\alpha}(\vec{\lambda})\right^2 J_{RC,\alpha}(\omega).
\label{eq: EFF-spec}
\eea
For notation convenience, in what follows we suppress the vector notation 
on $\vec{\lambda}$ and use identical RC parameters for the $w$ and $c$ baths,
$\Omega_{\alpha} = \Omega$, $\lambda_{\alpha} = \lambda$,
 as well as $\gamma_{\alpha} = \gamma$ for $\alpha=h,w,c$.
We now make several remarks pertaining to the effective QAR Hamiltonian Eq. (\ref{eq:EFF-QAR}):
%Furthermore, a schematic representation of the various mappings and transformations as well as the relationships between them is given in Fig. \ref{fig: mappingdiagram}. 

%(i)  

(i) The Effective QAR model is justified in Figs. \ref{S1} and \ref{S2}. First,
in Fig. \ref{fig:S1} we show the 6 lowest eigenenergies of the extended system Hamiltonian for $M=6$,
[Eq. (\ref{eq: HES})], presented as a function of the coupling strength $\lambda$. 
There is an energy separation of order $\Omega$ between the lowest three levels (corresponding to the
ground levels of the two RCs while the QAR is free to occupy any one of its three levels), 
and the next manifold of excited states.
Moreover, Fig. \ref{fig:S2} shows that the work bath, for example, mainly induces the transition between the 
states $|2({\lambda)}\rangle$ and $|3({\lambda})\rangle$, in correspondence with the original QAR model.
Similarly (not shown), the cold bath mainly induces the transition between the 
states $|1({\lambda})\rangle$ and $|2({\lambda})\rangle$, again 
in correspondence with the original QAR model.

(ii) Simulating transport characteristics with the EFF-QME is cheap and intuitive. 
We use the standard Redfield quantum master equation. This calculation is identical to what
is being done at weak coupling with the original QAR model of Eq. (\ref{eq: QAR Hamiltonian}); 
in the effective model, energies and coupling parameters implicitly depend on $\lambda$.

%The simple structure, mimicking the QAR Hamiltonian Eq. (\ref{eq: QAR Hamiltonian}) is further motivated in Fig. \ref{fig: fitting} where the matrix elements of $F_w(\lambda)\hat{S}_w$ are shown. Worthy of note, the only non-zero element is $\bra{b}\hat{V}_{ES,w}^D\ket{c}$, which connects the 2nd and 3rd excited state of the QAR while all other elements approximately vanish. 

(iii) 
The cooling performance of the effective QAR model can be readily obtained by repeating
standard derivations as in Ref. \cite{Correa14} (see also Appendix B in Ref. \cite{Kilgour2018}).
%Most crucially, the effective Hamiltonian, Eq. (\ref{eq:EFF-QAR}), allows for the identification of the cooling window simply by calculating the eigenvalues of the extended system Hamiltonian. 
Significantly, once building the effective model one does not need to perform any dynamical simulations to identify
the cooling window of the QAR in the strong coupling regime:
Using the Redfield quantum master equation, we write the cooling condition as
\bea
\frac{E_2{(\lambda)}-E_1{(\lambda)}}{E_3{(\lambda)}-E_1{(\lambda)}}
\leq\frac{\beta_h-\beta_w}{\beta_c-\beta_w},
\eea
which is analogous to the weak coupling result.
We demonstrate this point in Fig. \ref{fig:test1}.
In the presented parameters,  $\frac{\beta_h-\beta_w}{\beta_c-\beta_w} = 0.4$, which defines the cooling region.
We observe that for $\Delta =0.6$, the system enters the cooling region at $\lambda\approx 8$,
but leaves it at a stronger coupling when $\lambda\approx 11$ (see also Fig. 2(b) in the main text).
In contrast, at smaller value, $\Delta=0.2$, the system cools at weak-intermediate coupling, then stops cooling when $\lambda\approx 9$. This modification of the cooling condition with $\lambda$ corresponds to the reshaping of the
cooling window due to the renormalization of the energy levels 
with $\lambda$, as observed in Fig. 2 of the main text.
It should be noted that one can reach extreme situations when levels cross,
$E_1(\lambda) > E_2(\lambda)$, see Fig. \ref{fig:S1}. 
In this case (region ${\mathcal R}_4$ in Fig. 2(b) of the main text), 
the machine no longer functions as a QAR.
%In more details: In region ${\mathcal R}_4$ the renormalization effect is severe such that 
%allowed transitions are altered and a resonant condition can never be satisfied;
%the work bath couples to a transition of energy $E_3(\lambda)-E_2(\lambda)$, which becomes
%greater than the spacing $E_3(\lambda)-E_1(\lambda)$ controlled by the hot bath. 
%Thus, in this scenario heat is pumped in from the work bath towards the other reservoirs.

(iv) Failure of the EFF-QME approach to mimic the full RC-QME behavior 
allows us to pinpoint strong coupling effects
beyond parameter renormalization, specifically, the onset of heat leaks (heat transfer from the work bath 
towards the cold one).
We discuss heat leaks in Sec. \ref{sec: Leak}. 
%can be interpreted to switch between the work and the hot baths, i.e. $\hat{S}_w = \ket{1}\bra{3}$ and $\hat{S}_h = \ket{2}\bra{3}$, but with $T_w>T_h$. Such a thermal machine will never satisfy the cooling condition, and thus it ceases to act as a QAR. 
%In the main text, we refer to this regime as $\mathcal{R}_4$.

%=====================================================================================
% Figure S3 
\begin{figure}[hbt!] 
  \centering
 \includegraphics[width=0.5\linewidth]{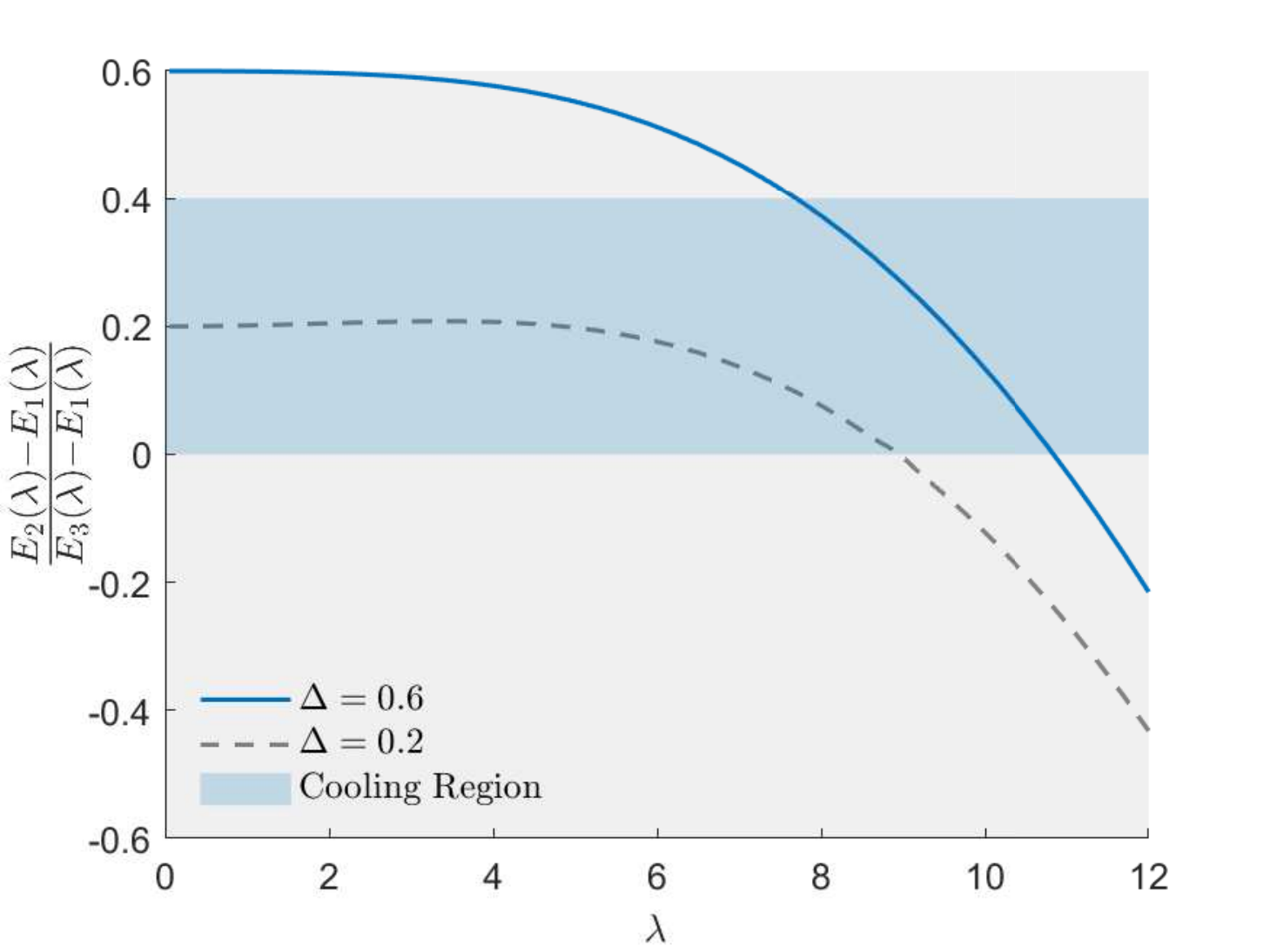}
 \caption{$\frac{E_2{(\lambda)}-E_1{(\lambda)}}{E_3{(\lambda)}-E_1{(\lambda)}}$ as a function of $\lambda$, illustrating the effects of parameter renormalization on the shape of the cooling window.
  %To satisfy the condition for cooling, the absolute value of this quantity must be less than $\frac{\beta_h-\beta_w}{\beta_c-\beta_w}$ corresponding to 0.4 for the parameters used in this work. 
  The region satisfying this condition is shaded in blue. In this figure, $\Delta = 0.6$ (\sblack) and $\Delta = 0.2$ (\dablack), all other parameters are identical to Fig. \ref{fig:S1} with $T_c = 0.25$, $T_h=0.5$, $T_w=1.5$.}
  \label{fig:test1}
\end{figure}
%=================================================================

\vspace{1cm}

\begin{figure}[hbt!]
\centering
\includegraphics[width=0.9\columnwidth]{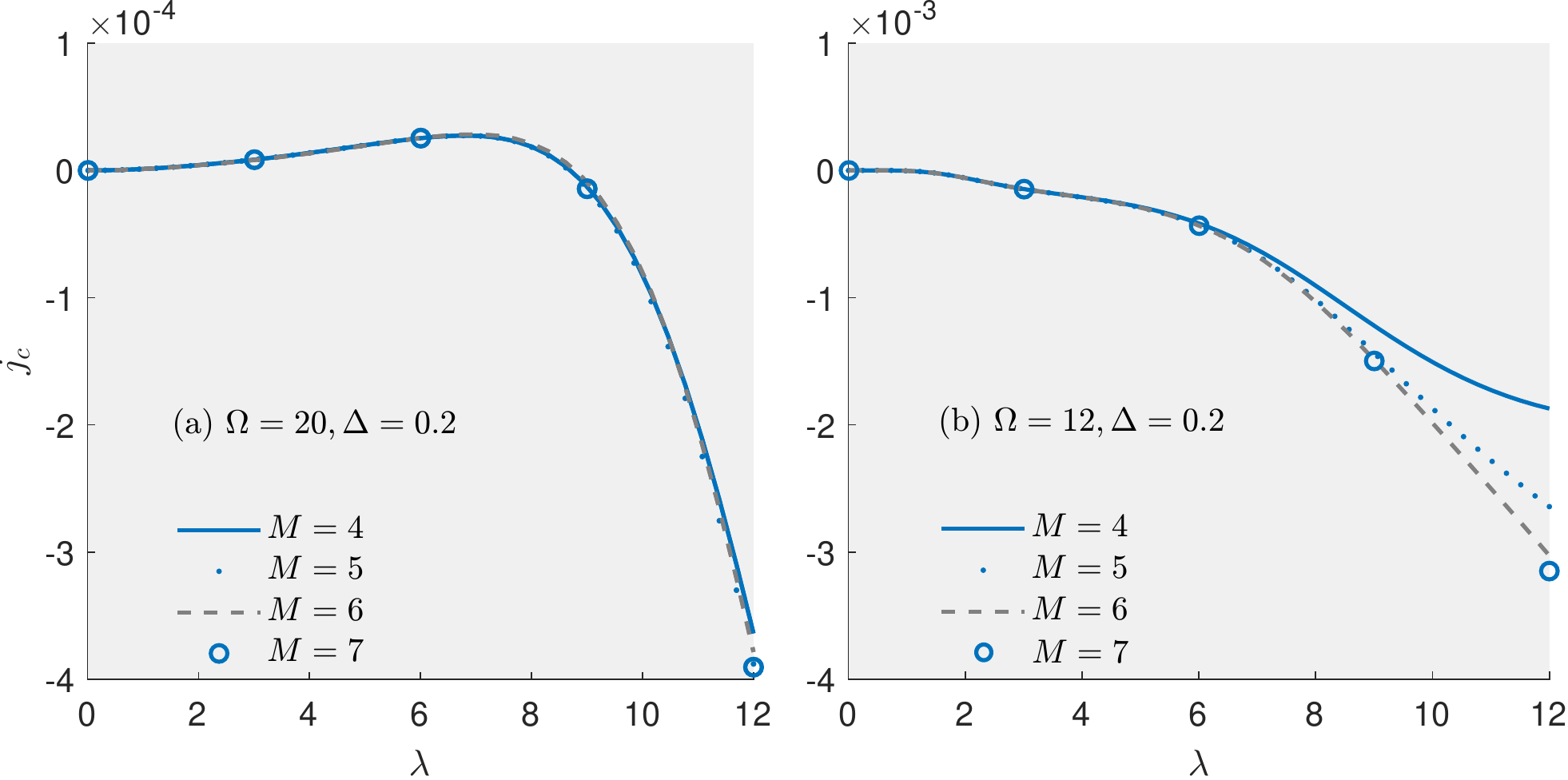}
\caption{Cooling current $j_c$ as a function of $\lambda$ truncating the RCs at
$M = 4$ (\sblack), $M = 5$ (\dblack), $M = 6$, (\dablack) and $M=7$ (hollow circles) with
(a) $\Omega = 20$, (b) $\Omega = 12$.
Other parameters are 
%identical to Fig. \ref{fig:S1} with 
$\gamma = 0.0071/\pi$, $\Lambda_c = \Lambda_h = \Lambda_w = 500$,  $T_c = 0.25$, $T_h=0.5$, $T_w=1.5$.
    %$\Delta = 0.2$, $T_c = 0.25$, $T_h=0.5$, $T_w=1.5$, $\gamma = 0.0071/ \pi$, $\Lambda = 500\pi$, $\epsilon_3-\epsilon_1=1$.
    }
    \label{fig: fitting}
\end{figure}

%=======================

%Section 3
\section{Convergence of RC-QME simulations}
\label{S4}

To describe the transport properties of QARs at strong coupling based 
on the RC-QME method, we truncate the reaction coordinate spectrum to include only 
the $M$ lowest levels from the harmonic oscillator manifold, see Sec. \ref{S2} and Eq. (\ref{eq: HES}). 
The number of levels required to reach convergence depends 
on the relative magnitudes of the RC frequency $\Omega$, the temperature of the 
environment $T_{\alpha}$ and the coupling strengths $\lambda$.
To converge, $M$ should be large enough such that
the probability of exciting the RC oscillators to that level is sufficiently small, thus of insignificant
contribution to transport.

In the main text, we utilized $M=6$ levels (unless otherwise stated). 
In Fig. \ref{fig: fitting}(a) we show that results were converged with $M$ already for $M=4$
 when  $\Omega =20$.
In contrast, for RCs with smaller $\Omega$,  Fig. \ref{fig: fitting}(b) shows
satisfying convergence only at $M=7$ once $\lambda>10$.

%================================================================
%The fullest description of the enlarged system requires incorporating all levels of the RC oscillators -- which is infinite and thus, intractable. Instead, we opt for a truncation procedure where only the first $M$-levels of the RC manifolds are kept.
%This procedure is physically grounded when the temperature is low enough, such that further excitations of the RC oscillators are made thermally improbable. More particularly, we justify this choice of $M$ (set to 6 throughout) by numerical convergence in Fig. \ref{fig: fitting}: for both $\Omega = 12$ and $\Omega = 20$, the result with $M = 6$ converges to that with $M = 7$. These imply that the result for $\Omega = 30$ also converges.
%======================================================
%=======================
% Figure S6

\begin{figure}[ht]
\includegraphics[width=0.9\columnwidth]{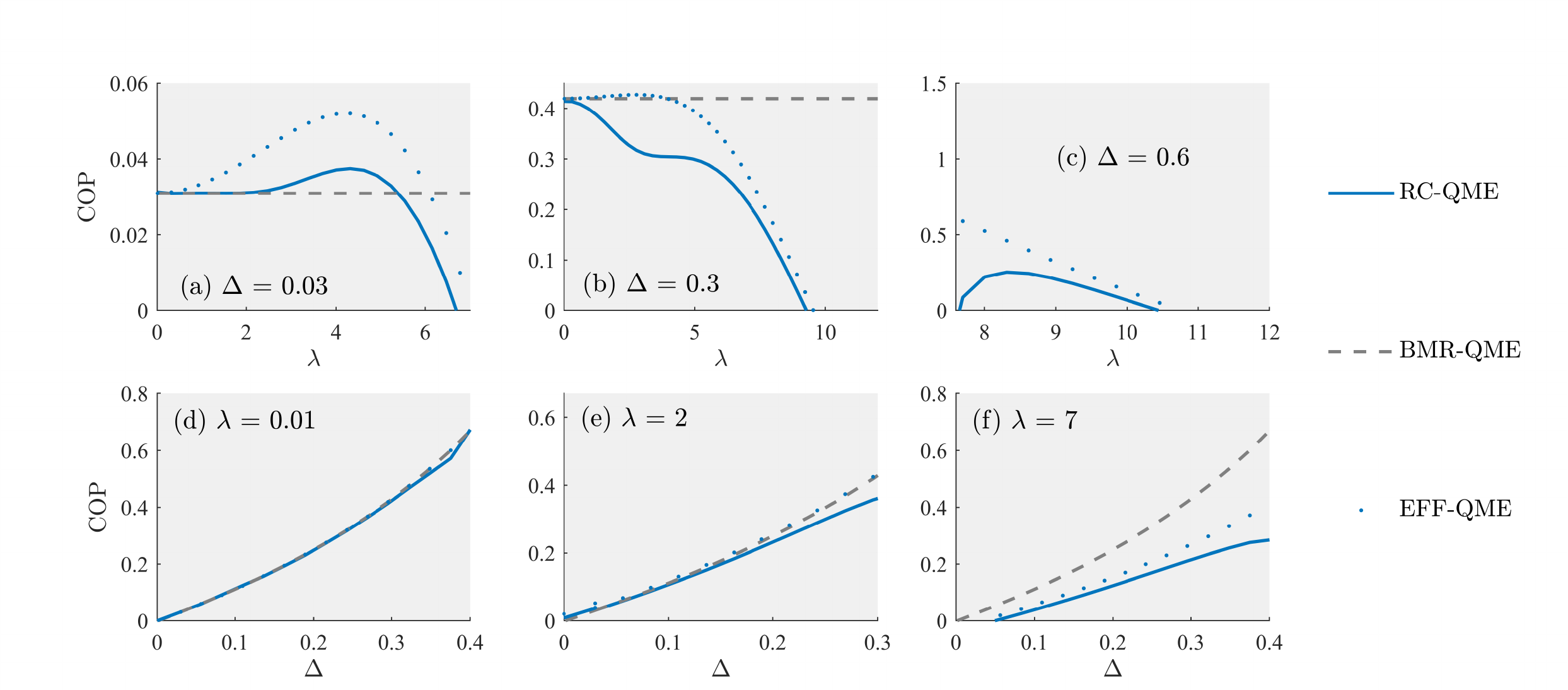}
\caption{Coefficient of performance (COP) as a function of
(a) $\lambda$ with $\Delta = 0.03$,
(b) $\lambda$ with $\Delta = 0.3$,
(c) $\lambda$ with $\Delta = 0.6$,
(d) $\Delta$ with $\lambda = 0.01$,
(e) $\Delta$ with $\lambda = 2$, and
(f) $\Delta$ with $\lambda = 7$.
Calculations were done with the RC-QME (\sblack),
the BMR-QME (\dblack) and the EFF-QME (\dablack).
Parameters in this plot are identical to Fig. \ref{fig: fitting} with $\Omega=20$, $M=6$. }
%with $\gamma = 0.0071/\pi$, $\Lambda_c = \Lambda_h = \Lambda_w = 500$.
 %$\Omega=20$, $T_c = 0.25$, $T_h=0.5$, $T_w=1.5$, $\gamma = 0.0071/ \pi$, $\Lambda = 500\pi$, $\epsilon_3-\epsilon_1=1$.}}
\label{fig: COP}
\end{figure}
% DDD fix fugyre QME and caption
%=======================
% Figure S7
\begin{figure}[ht]
\centering
\includegraphics[width=1\columnwidth]{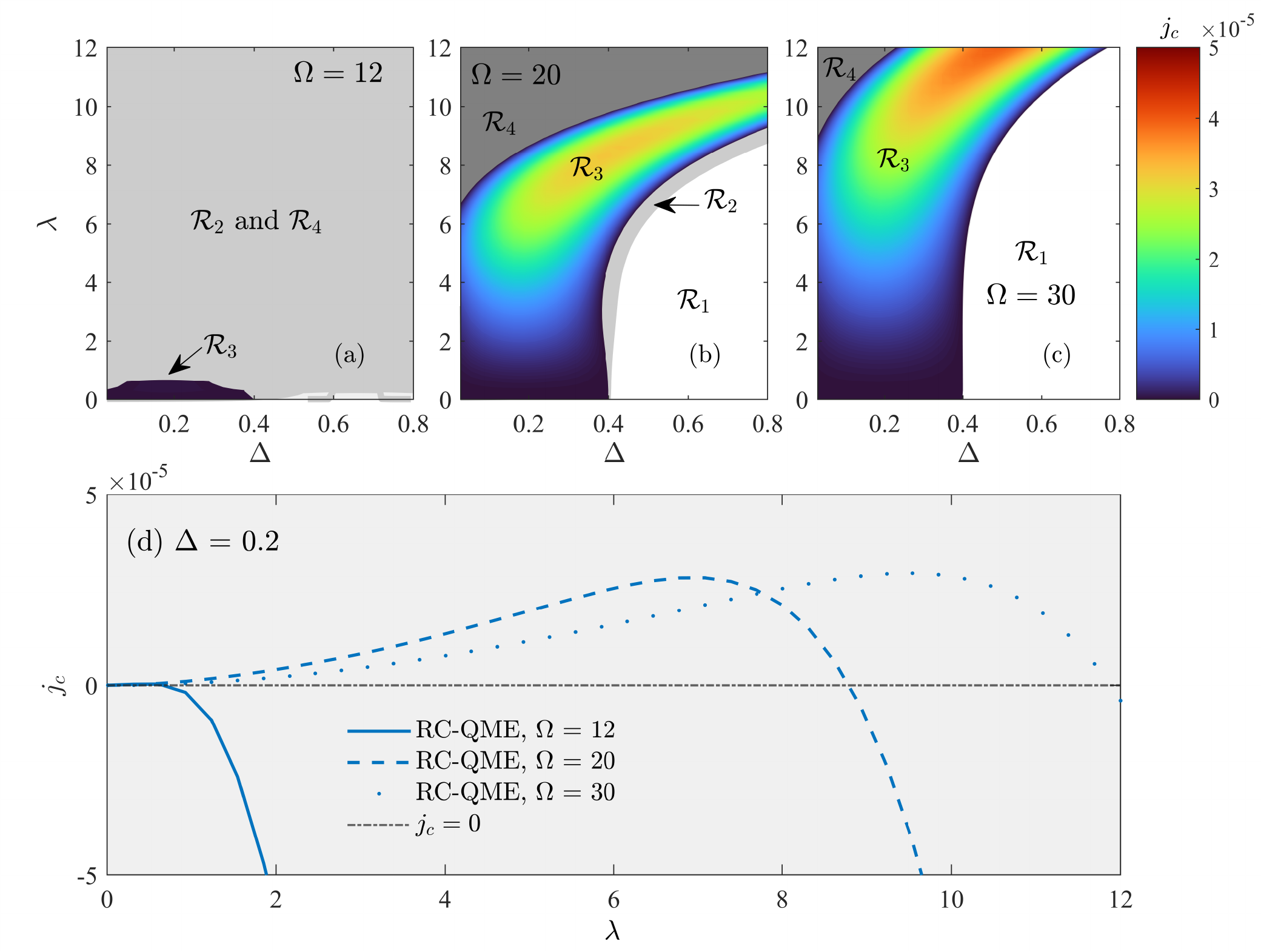}
\caption{(a)-(c) Contour maps of the cooling current
$j_c$ as a function of the coupling strength $\lambda$ and the energy gap $\Delta$
using the the RC-QME method with $M = 6$ for
(a) $\Omega = 12$,
(b) $\Omega = 20$,
(c) $\Omega = 30$.
(d) We present slices of the contour maps, $j_c$ as a function of $\lambda$ with
at $\Delta=0.2$, and $\Omega = 20$ (dashed), $\Omega = 12$ (full), $\Omega = 30$ (dotted).
Other parameters are the same as in Fig. \ref{fig: fitting}.}
\label{fig: OM}
\end{figure}

%====================

%===========================================
\section{Additional simulations}
\label{S5}

\subsection{Coefficient of performance}
We present additional simulations of the coefficient of performance (COP) for the QAR, $\text{COP} \equiv \frac{j_c}{j_w}$,
 in the strong coupling regime.
Fig. \ref{fig: COP} displays the COP  from the nonperturbative RC-QME method,
compared to the BMR-QME technique that is valid only 
at weak system-reservoir coupling, and the EFF-QME approach. 

If the system is able to cool at weak coupling, we find that strong coupling is typically detrimental to the performance of the QAR, see
 Fig. \ref{fig: COP}(a)-(b) where the COP is typically suppressed with increasing $\lambda$.
Note that in Fig. \ref{fig: COP}(a) when there is a small enhancement of the COP for intermediate $\lambda$,
the actual cooling current is very small since $\Delta$ is small. 
In  contrast, in Fig. \ref{fig: COP}(c) the BMR-QME method does not predict any cooling. Here,
as we increase $\lambda$ the RC-QME method reveals that we enter the cooling window with a finite COP,
 increasing it until a certain threshold where it starts to be suppressed.
%The impact of parameter renormalization is prominent in  Fig. \ref{fig: COP}(c) 
%At large coupling, $\lambda \approx 8$, 
%parameter renormalization of the QAR levels allows the system to satisfy the cooling condition, 
%entering the cooling window. However, as we further increase couplings to
%$\lambda \approx 10$, the cooling condition is no long fulfilled and the system exits the cooling window. 
%
Fig. \ref{fig: COP}(d)-(f) show that as we increase $\Delta$ for a given coupling strength, we monotonically
enhance the COP.

% DDD we need to compare to the small lambda behavior, not to the BMR-QME, which is invalid!
%This can be seen in Fig \ref{fig: COP}(b)-(f) where the COP as calculated 
%by the weak coupling technique (BMR-QME) surpasses the values calculated by the strong coupling 
%RC-QME method. Fig. \ref{fig: COP} (a) is an exception, where, for very small splitting $\Delta$, the COP as calculated by the RC-QME is marginally larger than weak coupling methods. Despite this, it is still not useful. This regime provides a tiny cooling power (see Fig. \ref{fig: OM}) such that strong coupling offers no functional benefit from the point of view of performance. 

Fig. \ref{fig: COP}(a)-(f) demonstrate that the COP as calculated via the EFF-QME method 
overestimates the RC-QME technique, though it captures central trends with $\lambda$ and $\Delta$.
This is to be expected since the EFF-QME method does not take into account
the detrimental roles of inter-bath heat flow and internal dissipation, both 
contributing to reduce the COP for the RC-QME.

%Note, the result we show for the BMR-QME is misleading: COP is still positive despite being outside the cooling window.
%maybe don't need this...
%This last point reflects a key, and central aspect of this work: Strong coupling provides an alternative parameter regime for cooling, by shifting the cooling window due to the renormalization of QAR parameters.

%====================
\subsection{Inter-bath (leakage) heat transport}
\label{sec: Leak}
% DDD exmplify: heat flow from an

So far, we discussed the role of parameter renormalization as a central aspect of strong system-bath
coupling in QARs.  This aspect was rationalized with the effective description, 
Eq. (\ref{eq:EFF-QAR}), valid if $\Omega$, the frequency of the reaction coordinate was made large compared to the temperature of the baths and $\lambda$.
Since in the effective 3-level QAR description only parameter renormalizations is captured,
 deviations of its predictions from the RC-QME, as observed in Fig. \ref{fig: COP},
are to be associated with additional strong coupling mechanisms.  %which are beyond level renormalization. As we will point out shortly, these strong coupling mechanisms are detrimentally affecting the QAR.

Indeed, another novel impact of strong system bath coupling is the opening of a direct
heat flow pathway from the work bath towards the cold one. 
In this process, an excited RC from the work bath transfers its energy to the RC of the cold bath, even
with little change to the population of the QAR levels.
In  Fig. \ref{fig: OM} we show that
by tuning $\Omega$, we  control the extent of such a direct inter-bath transport pathway.
%and how it negatively affects the QAR cooling window and cooling power.
To understand these results,  we label four distinct regions in the contour plot: 

In region $\mathcal{R}_3$,   $j_c > 0$ and $j_w > 0$, and we are thus in the cooling window
with heat extracted from the cold bath, assisted by the work bath, and dumped into the hot one.
In regions $\mathcal{R}_2$ and $\mathcal{R}_4$, we do not observe cooling as 
$j_c < 0$. Notably, $j_w > 0$ in these regions thus heat flows from the work bath into the cold one. This direct heat flow from the
work bath to the cold one, which cannot take place at weak coupling in the 3-level QAR model, is referred to as inter-bath heat transport,
and it is responsible of heat leaks in the present case.
We distinguish between $\mathcal{R}_2$ and $\mathcal{R}_4$ below.
The system does not cool in $\mathcal{R}_1$ with both $j_c < 0$ and $j_w < 0$, signifying that heat flows from the hot bath towards the other two reservoirs.

We now discuss Fig. \ref{fig: OM}. First, in panel (a) $\Omega$ is relatively small ($\Omega=12$)
and inter-bath leakage is a primary effect of strong coupling:
The cooling window is severely decreased due to inter-bath transport effects.
In Figure \ref{fig: OM}(b), we display the same results as in Fig. 2(b) from the main text.
%highlighting when inter-bath transport is a secondary effect of strong coupling, 
%having both this effect and level-renormalization present. 
In this case of $\Omega=20$, we clearly observe four different regions: $\mathcal{R}_3$ is the cooling window, 
which is significantly broader than the cooling window found in Fig. \ref{fig: OM}(a).
%due to the inter-bath transport not being a major contribution.
Region $\mathcal{R}_1$ corresponds to the  ``regular" no-cooling zone observed in the weak coupling limit.
Next, $\mathcal{R}_2$ illustrates where the effect of inter-bath transport takes over, 
resulting in a smaller cooling window. 
Finally, $\mathcal{R}_4$ represents a region where there is no cooling due to the breakdown 
of the ``correct" level scheme of the  QAR (see Fig. \ref{S1}); 
in this region, $E_2(\lambda)<E_1(\lambda)$, 
%{E_3{(\lambda)}-E_1{(\lambda)}}\leq\frac{\beta_h-\beta_w}{\beta_c-\beta_w}$ does not hold, 
and so we are outside the cooling window. % but for different reasons than for $\mathcal{R}_2$. 

Fig. \ref{fig: OM}(c) depicts the case of even larger RC frequency, $\Omega = 30$, such that
  excited states of the RC are negligibly populated. As a result, direct inter-bath transitions do not 
contribute and region $\mathcal{R}_2$ is absent from the plot.

\end{widetext}